\documentclass[twocolumn]{emulateapj}
\usepackage{apjfonts}

\usepackage{natbib}
\usepackage{epsfig}
\usepackage{rotating}

\newcommand{\be}{\begin{equation}}
\newcommand{\ee}{\end{equation}}

\newcommand{\kms}{\mbox{km\,\ensuremath{\rm{s}^{-1}}}}

\shortauthors{Cordiner et al.}

\begin{document}

\submitted{Accepted for publication in ApJ, February 2017}

\title{ALMA mapping of rapid gas and dust variations in comet C/2012 S1 (ISON): new insights into the origin of cometary HNC}

\author{M. A. Cordiner\altaffilmark{1,2}, J. Boissier\altaffilmark{3}, S. B. Charnley\altaffilmark{1}, A. J. Remijan\altaffilmark{4}, M. J. Mumma\altaffilmark{1}, G. Villanueva\altaffilmark{1,2}, D. C. Lis\altaffilmark{5}, S. N. Milam\altaffilmark{1}, L. Paganini\altaffilmark{1,2}, J. Crovisier\altaffilmark{6}, D. Bockel{\'e}e-Morvan\altaffilmark{6}, Y.-J. Kuan\altaffilmark{7,8}, N. Biver\altaffilmark{6}, I. M. Coulson\altaffilmark{9}}


\altaffiltext{1}{NASA Goddard Space Flight Center, 8800 Greenbelt Road, Greenbelt, MD 20771, USA.}
\email{martin.cordiner@nasa.gov}
\altaffiltext{2}{Department of Physics, Catholic University of America, Washington, DC 20064, USA.}
\altaffiltext{3}{IRAM, 300 Rue de la Piscine, 38406 Saint Martin d'Heres, France.}
\altaffiltext{4}{National Radio Astronomy Observatory, Charlottesville, VA 22903, USA.}
\altaffiltext{5}{LERMA, Observatoire de Paris, PSL Research University, CNRS, Sorbonne Universit{\'e}s, UPMC Univ. Paris 06, F-75014, Paris, France.}
\altaffiltext{6}{LESIA, Observatoire de Paris, CNRS, UPMC, Universit{\'e} Paris-Diderot, 5 place Jules Janssen, 92195 Meudon, France.}
\altaffiltext{7}{National Taiwan Normal University, Taipei 116, Taiwan, ROC.}
\altaffiltext{8}{Institute of Astronomy and Astrophysics, Academia Sinica, Taipei 106, Taiwan, ROC.}
\altaffiltext{9}{East Asian Observatory, Hilo, HI 96720, USA.}

\begin{abstract}

Observations of the sungrazing comet C/2012 S1 (ISON) were carried out using the Atacama Large Millimeter/submillimeter Array (ALMA) at a heliocentric distance of 0.58-0.54~AU (pre-perihelion) on 2013 November 16-17. Temporally resolved measurements of the coma distributions of HNC, CH$_3$OH, H$_2$CO and dust were obtained over the course of about an hour on each day.  During the period UT 10:10-11:00 on Nov. 16, the comet displayed a remarkable drop in activity, manifested as a $>42$\% decline in the molecular line and continuum fluxes. The H$_2$CO observations are consistent with an abrupt, $\approx50$\% reduction in the cometary gas production rate soon after the start of our observations. On Nov. 17, the total { observed} fluxes remained relatively constant during a similar period, but strong variations in the morphology of the HNC distribution were detected as a function of time, indicative of a clumpy, intermittent outflow for this species. Our observations suggest that at least part of the detected HNC originated from degradation of nitrogen-rich organic refractory material, released intermittently from confined regions of the nucleus. By contrast, the distributions of CH$_3$OH and H$_2$CO during the Nov. 17 observations were relatively uniform, consistent with isotropic outflow and stable activity levels for these species. These results highlight a large degree of variability in the production of gas and dust from comet ISON during its pre-perihelion outburst, consistent with repeated disruption of the nucleus interspersed with periods of relative quiescence. 

\end{abstract}


\section{Introduction}

Sungrazing comet C/2012 S1 (hereafter referred to as comet ISON) made its first (and final) passage through the inner Solar System in 2013. It originated in the Oort Cloud at a distance $\gtrsim20,000$~AU, and its orbit passed within 0.013~AU of the Sun at perihelion on 2013 November 28. By the time comet ISON reached perihelion, its ($\sim300$~m diameter) nucleus had disintegrated \citep{kni14,lam14,kea16}, leaving a trail of dust and debris in its wake.

Between 2013 November 12-18, comet ISON's activity increased dramatically (see for example, \citealt{cro13}), as it traversed a heliocentric distance $r_H=0.8$-0.5~AU. During this period of outburst, the comet's visual brightness and water production rate increased by over an order of magnitude, and strong brightness fluctuations (on the order of a factor of two) began to occur over timescales of less than a day \citep{sek14,com14}. This event has been associated with fragmentation of the cometary nucleus and consequent increase in active surface area. The observation of coma wing features in the optical on November 14, 16, and 18 by \citet{boe13} may be taken as further evidence for the release of nucleus fragments. \citet{agu14} observed a dramatic increase in HCN production on Nov. 14, followed by a remarkable daily variability using the IRAM 30-m telescope during the period November 13-16, immediately preceding our ALMA observations.

Measurements of the spatial distributions of HCN, HNC and H$_2$CO gas as well as thermal dust emission from the inner coma of comet ISON were presented by \citet{cor14}, based on observations using the Atacama Large Millimeter/submillimeter Array on 2013 November 17, when the comet's active area reached a peak \citep{com14}. The HCN and H$_2$CO distributions were found to be quite symmetric, consistent with predominantly isotropic release of these molecules from within a few hundred kilometers of the nucleus. The HNC distribution, on the other hand, showed surprising collimated/clumpy features, suggesting release of this molecule within jets, streams or clumps. Distributed/extended coma sources for HNC and H$_2$CO were definitively identified, but the detailed production mechanisms for these molecules are still not well understood (see also \citealt{cot08}).

In this article, we use temporally-resolved ALMA observations to examine the detailed spatial distribution and time variability of emission from dust and molecules in comet ISON on 2013 November 16-17. The fine ($\approx 17$~min) time cadence of our molecular mapping provides powerful new insights into the short timescale variability of this comet during outburst, and helps elucidate the nature and origin of the anisotropic HNC features identified by \citet{cor14}.
 
\section{Observations}
\label{sec:obs}

\begin{table}
\centering
\caption{Summary of Observations \label{tab:obs}}
\begin{tabular}{ccll}
\hline\hline
Date&UT Time&Object&Scan\\
\hline
2013-11-16&10:03:47 - 10:04:49&3C\,279&1\\
          &10:05:44 - 10:08:21&Pallas&2\\
          &10:08:44 - 10:09:46&3C\,279&3\\
          &10:10:27 - 10:17:19&ISON&4\\
          &10:17:30 - 10:18:32&3C\,279&5\\
          &10:19:02 - 10:25:53&ISON&6\\
          &10:26:04 - 10:27:05&3C\,279&7\\
          &10:27:36 - 10:34:28&ISON&8\\
          &10:34:45 - 10:35:47&3C\,279&9\\
          &10:36:17 - 10:43:09&ISON&10\\
          &10:43:19 - 10:44:21&3C\,279&11\\
          &10:45:02 - 10:51:53&ISON&12\\
          &10:52:03 - 10:53:06&3C\,279&13\\
          &10:53:36 - 11:00:27&ISON&14\\
          &11:00:37 - 11:01:39&3C\,279&15\\[2mm]
2013-11-17&12:23:25 - 12:24:27&3C\,279&1\\
          &12:25:16 - 12:27:53&Titan&2\\
          &12:28:08 - 12:29:10&3C\,279&3\\
          &12:29:54 - 12:36:46&ISON&4\\
          &12:36:57 - 12:37:59&3C\,279&5\\
          &12:38:30 - 12:45:21&ISON&6\\
          &12:45:33 - 12:46:35&3C\,279&7\\
          &12:47:06 - 12:53:58&ISON&8\\
          &12:54:17 - 12:55:19&3C\,279&9\\
          &12:55:50 - 13:02:42&ISON&10\\
          &13:02:53 - 13:03:55&3C\,279&11\\
          &13:04:41 - 13:11:32&ISON&12\\
          &13:11:43 - 13:12:45&3C\,279&13\\
          &13:13:16 - 13:20:08&ISON&14\\
          &13:20:19 - 13:21:21&3C\,279&15\\
\hline
\end{tabular}
\end{table}

Observations were made in Cycle 1 Early Science mode using the (dual sideband) ALMA band 7 receiver, covering frequencies 350.3-352.3~GHz (lower sideband) and 362.2-364.1~GHz (upper sideband). Comet ISON was observed pre-perihelion on UT 2013-11-16 at 10:10-11:00 (at a heliocentric distance $r_H=0.58$~AU and geocentric distance $\Delta=0.89$~AU), and again on UT 2013-11-17 at 12:30-13:20 ($r_H=0.54$~AU, $\Delta=0.88$~AU). The illumination (sun-comet-observer) phase angle was $82^{\circ}-85^{\circ}$. The comet's position was tracked using JPL Horizons ephemeris solution \#45 and the coordinates of the ALMA phase center were updated in real time to account for the comet's rapid non-sidereal motion. A summary of these observations, which constitute a subset of the data acquired during our full ALMA/ISON observing campaign, is given in Table \ref{tab:obs}. 

The ALMA correlator was configured with four spectral windows to simultaneously observe all the lines of interest plus continuum on each day (see Table \ref{tab:lines}). Observations of the bright quasar 3C\,279 were used to calibrate the instrumental bandpass, phase and gain. For calibration of the flux scale, observations of the Solar System bodies Pallas and Titan were used, resulting in absolute fluxes accurate to within about 15\%. Using $28\times12$-m antennae (with baseline lengths in the range 17.3-1284 m), the angular resolution was $0.28''\times0.44''$. The correlator channel spacing of 244~kHz resulted in a spectral resolution of 488~kHz (about 0.4~\kms) after Hanning smoothing in the time domain. Weather conditions were excellent for all observations, with precipitable water vapor column at zenith of less than 0.57~mm. The RMS of the phase variation was measured for all visits to the phase calibrator and for all baseline pairs, with median RMS values of $15.8^{\circ}$ on November 16 and $6.4^{\circ}$ on November 17. These phase variations tracked and corrected accordingly for each antenna.

The data were flagged, calibrated and imaged using standard routines in CASA version 4.1.0 \citep{mcm07}. No significant flux was detected for baselines $\gtrsim400$~m, so the most distant antennae (DV07, DV19 and DV25; $>500$~m from the array center) were excluded during imaging to provide a more uniform Fourier sampling pattern (in the $uv$ plane), resulting in a corresponding improvement in image quality at the expense of a small degradation in angular resolution (to $0.40''\times0.54''$). Deconvolution of the point-spread function (PSF) was performed using the H{\"o}gbom algorithm, with natural visibility weighting and a flux threshold of twice the expected RMS noise per channel in each image.  Initially, each science scan (see Table \ref{tab:obs}) was cleaned and imaged individually. However, to improve the signal-to-noise ratio for the purpose of the present study, the scans taken on each day were combined into pairs, by averaging scans 4\&6, scans 8\&10 and scans 12\&14 into three discrete observation periods (hereafter referred to as periods 1, 2 and 3). Each period had a total on-source integration time of 13.7 min and the time interval between periods was approx. 17 min.

Comet ISON's 0.8~mm continuum was imaged by averaging the observed data over all line-free channels. The continuum peak was found to be offset by $5.7''$ from the predicted ephemeris position at the start of observations on Nov 16, and by $6.6''$ on Nov 17. The direction of offset was the same on both days: $75^{\circ}$ clockwise from celestial north, compared with the $71^{\circ}$ position angle of the Sun-comet vector. A similar offset was measured by optical observers, explainable as a result of non-gravitational acceleration of the comet due to preferential outgassing in the sunward-facing hemisphere \citep{sek14}. To maintain accurate flux measurements across our field of view, the interferometric images were corrected for the Gaussian spatial response of the ALMA primary beam (half-power beam-width = $16''$ at 363~GHz). Finally, the images were transformed from celestial coordinates to cometocentric (sky-projected) spatial distances, the origin of which was determined as the mean location of the continuum flux peak on each day. The coordinate axes are aligned with the equatorial system (celestial north is up).

\section{Results}
\label{results}

\begin{figure*}
\centering
UT 2013-11-16\ \ 10:18
\hspace{3.2cm}
UT 2013-11-16\ \ 10:35
\hspace{3.2cm}
UT 2013-11-16\ \ 10:53
\includegraphics[width=0.32\textwidth]{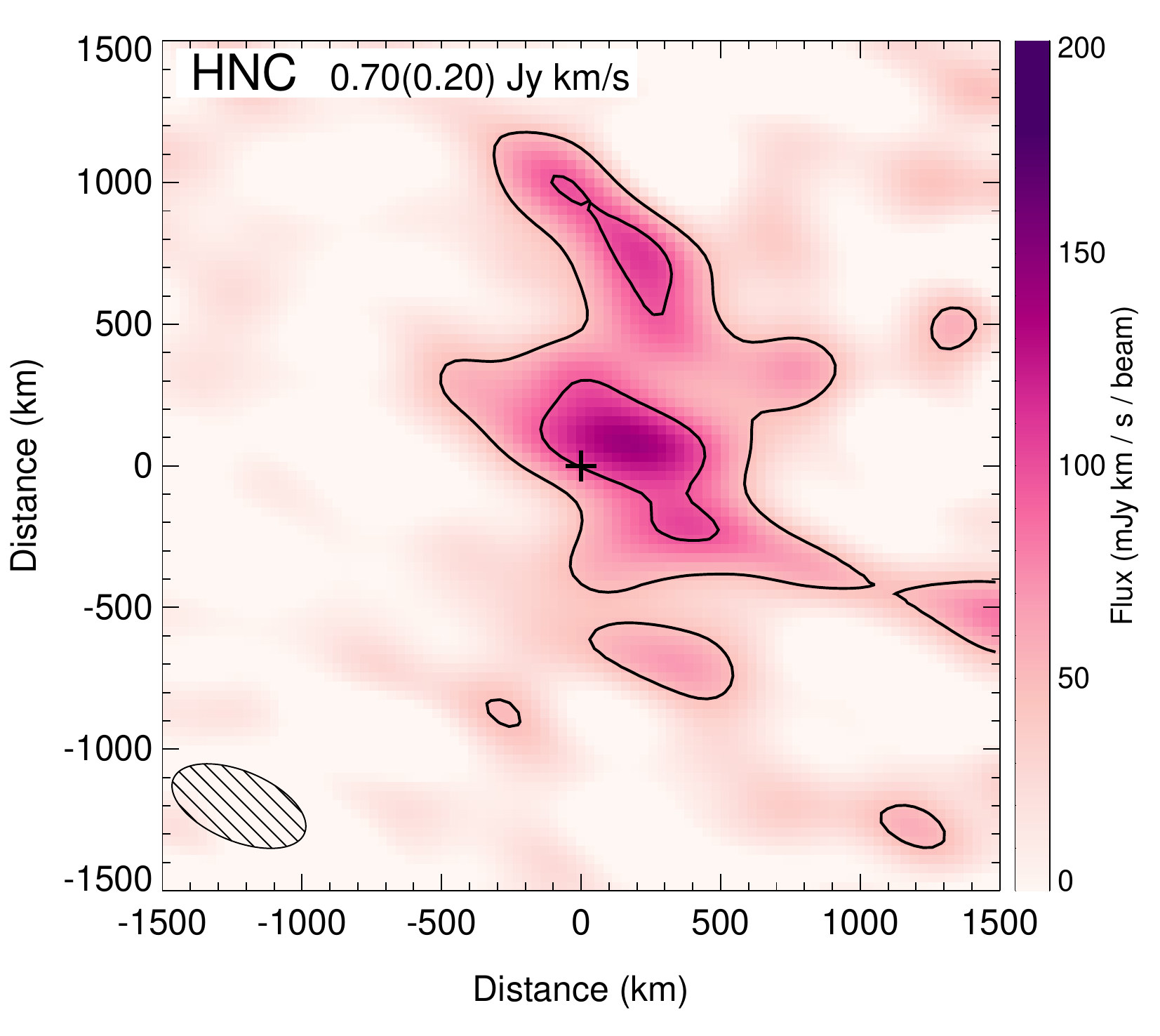}
\includegraphics[width=0.32\textwidth]{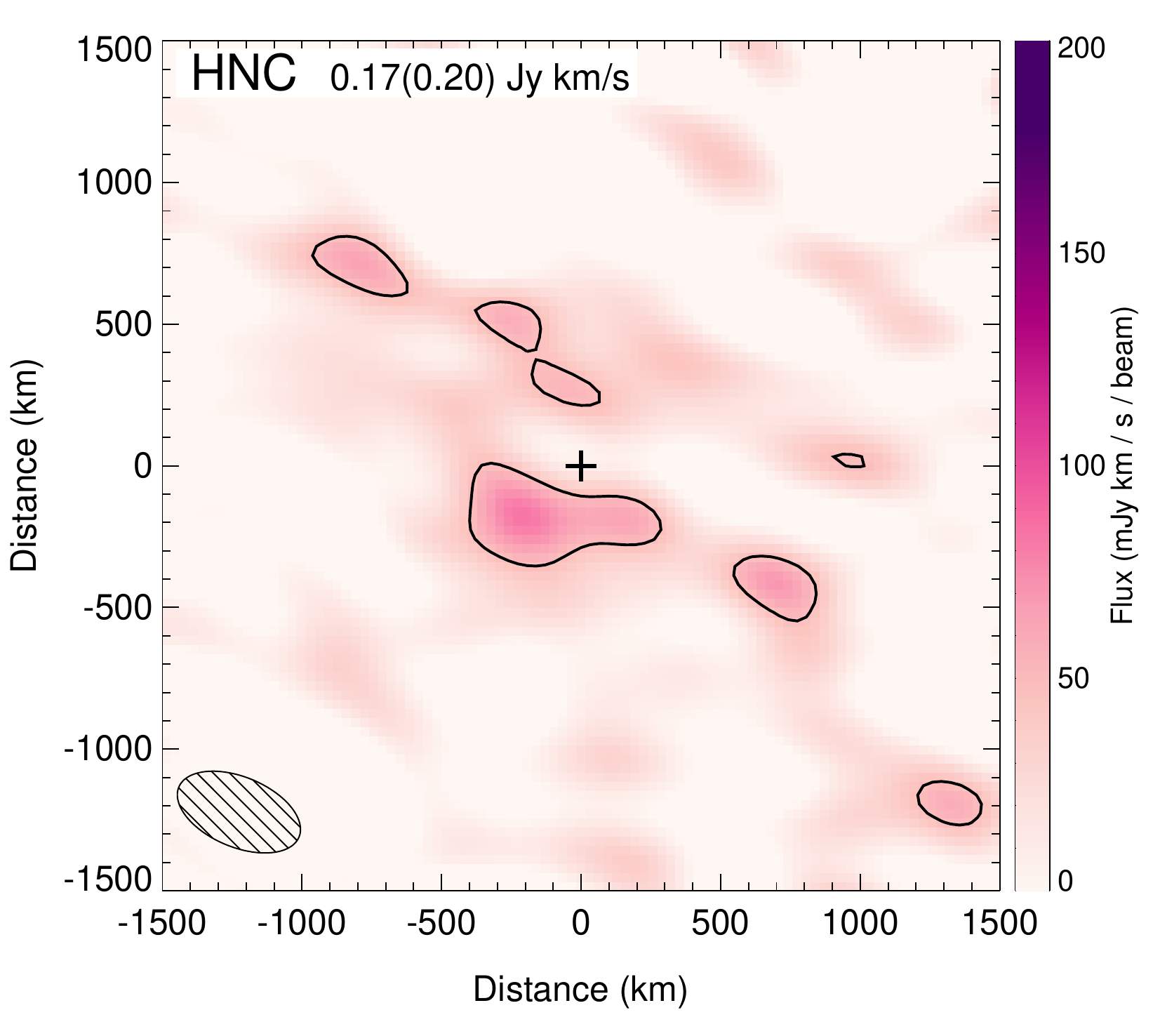}
\includegraphics[width=0.32\textwidth]{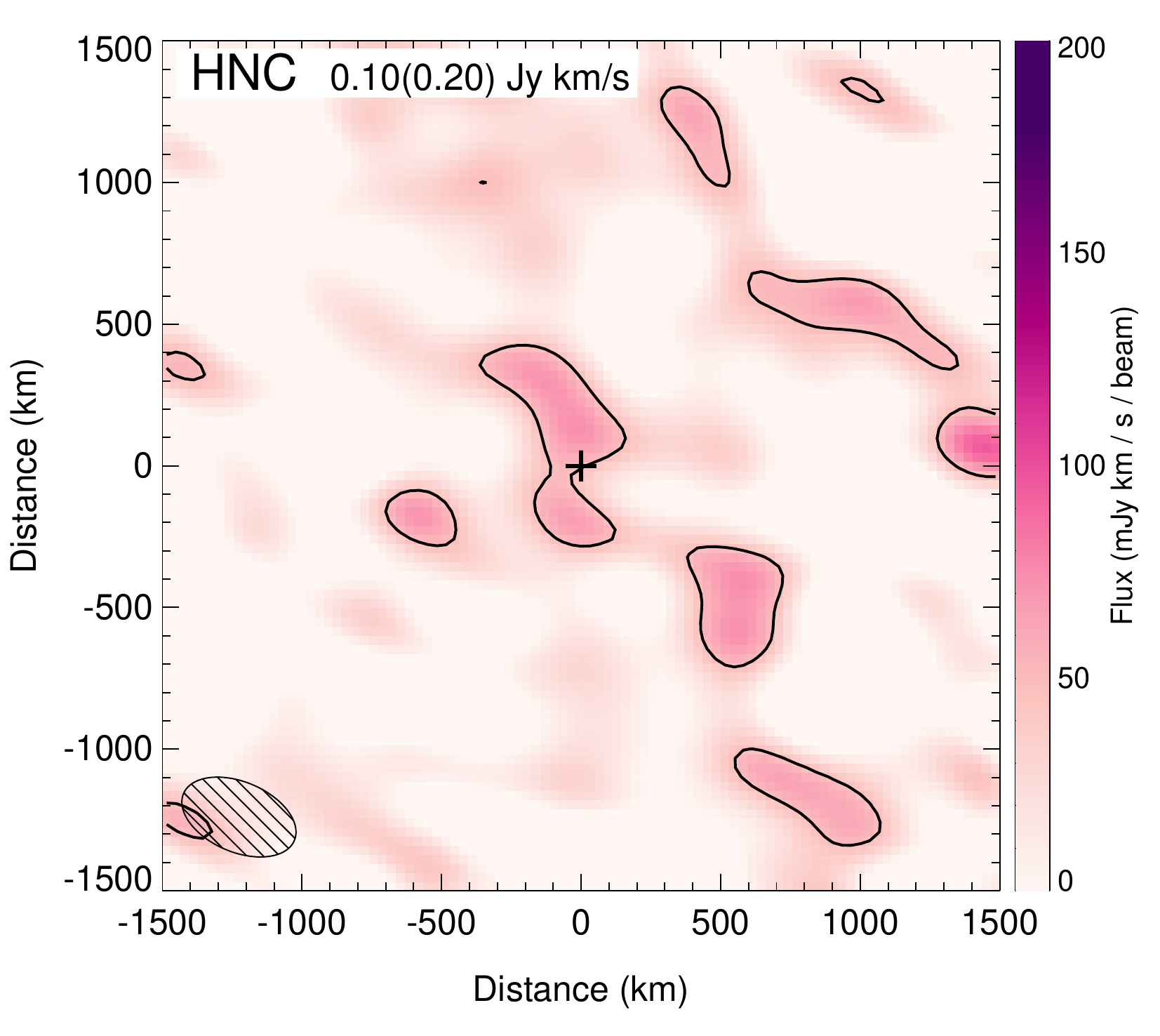}\\
\includegraphics[width=0.32\textwidth]{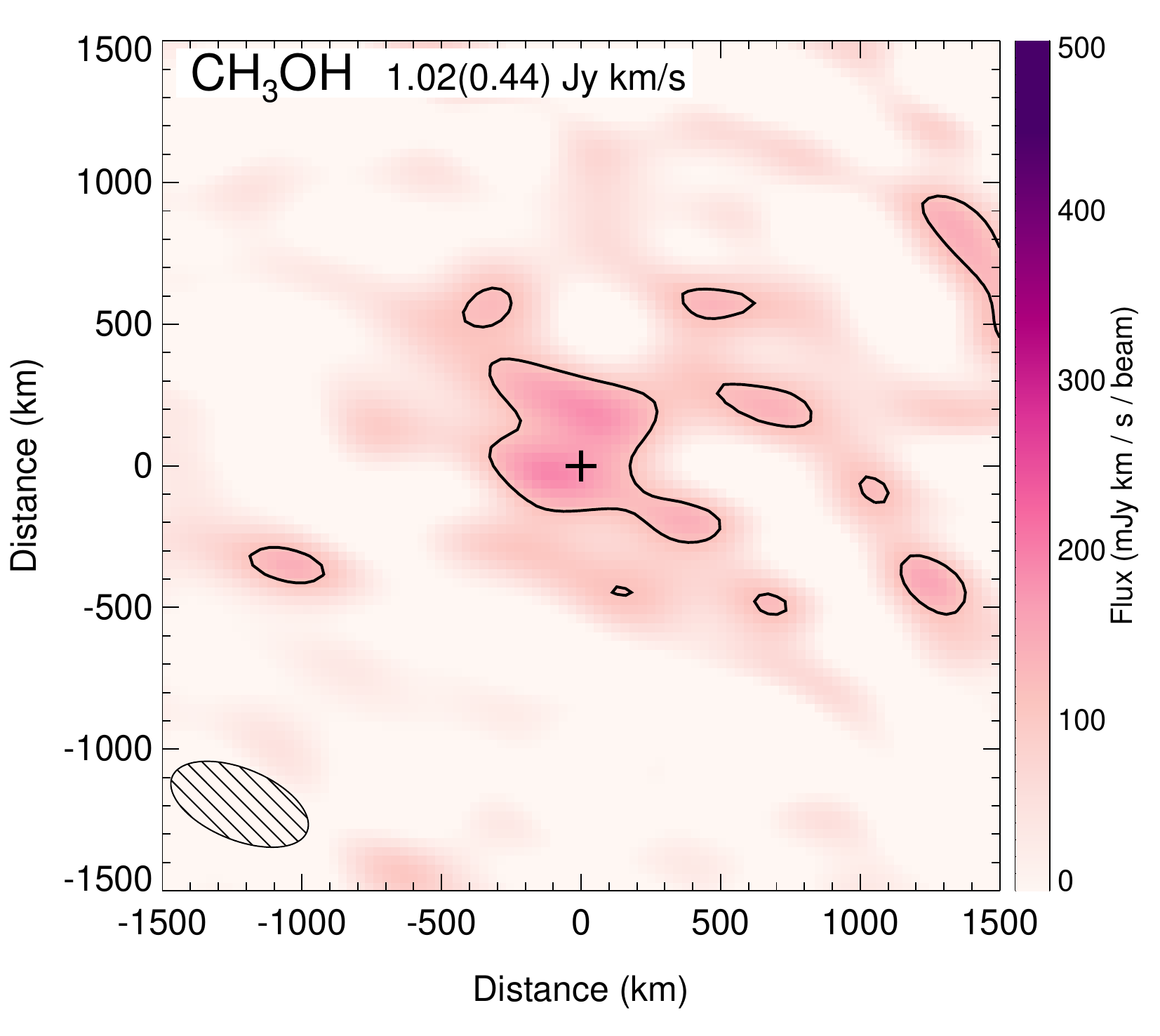}
\includegraphics[width=0.32\textwidth]{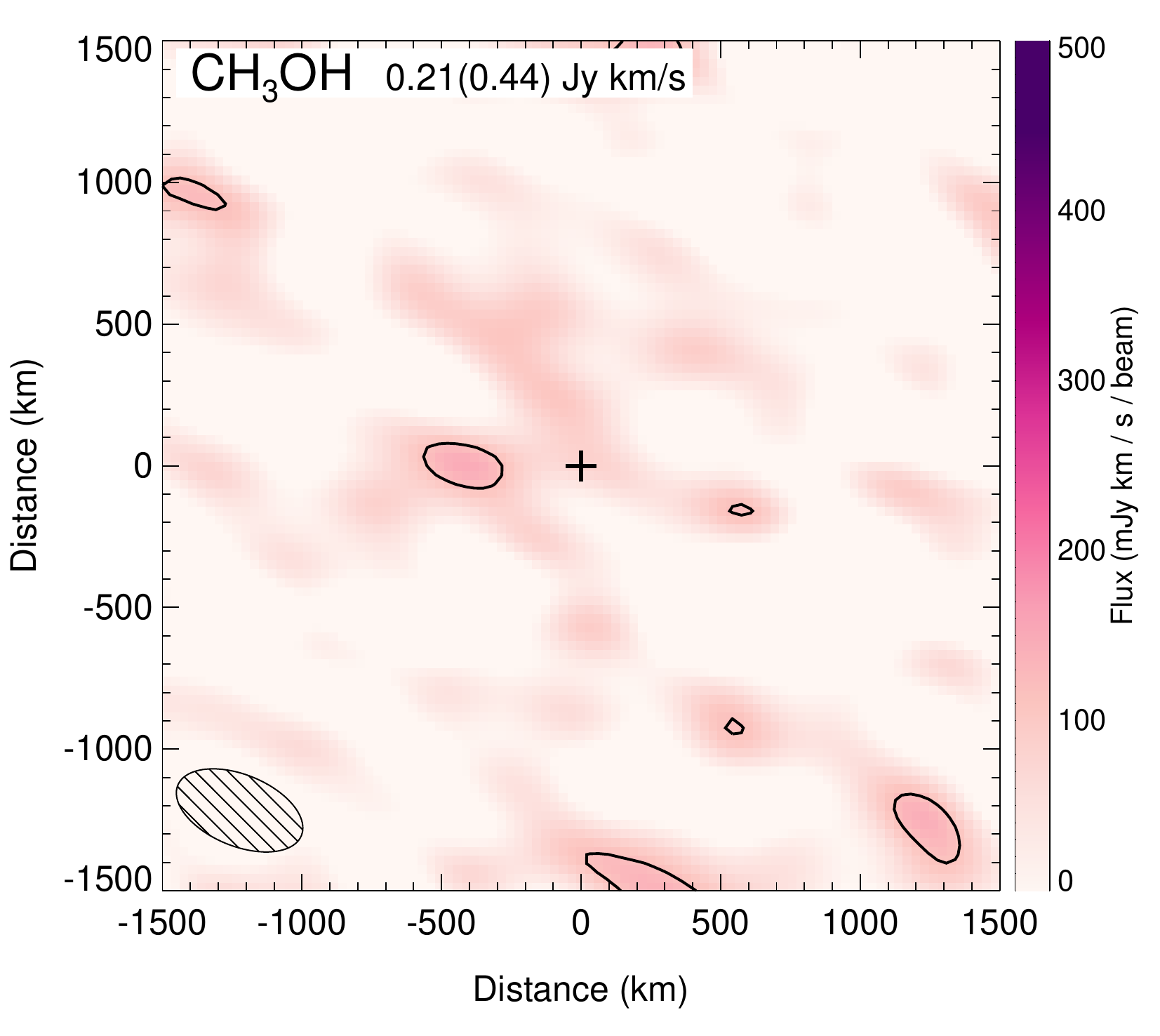}
\includegraphics[width=0.32\textwidth]{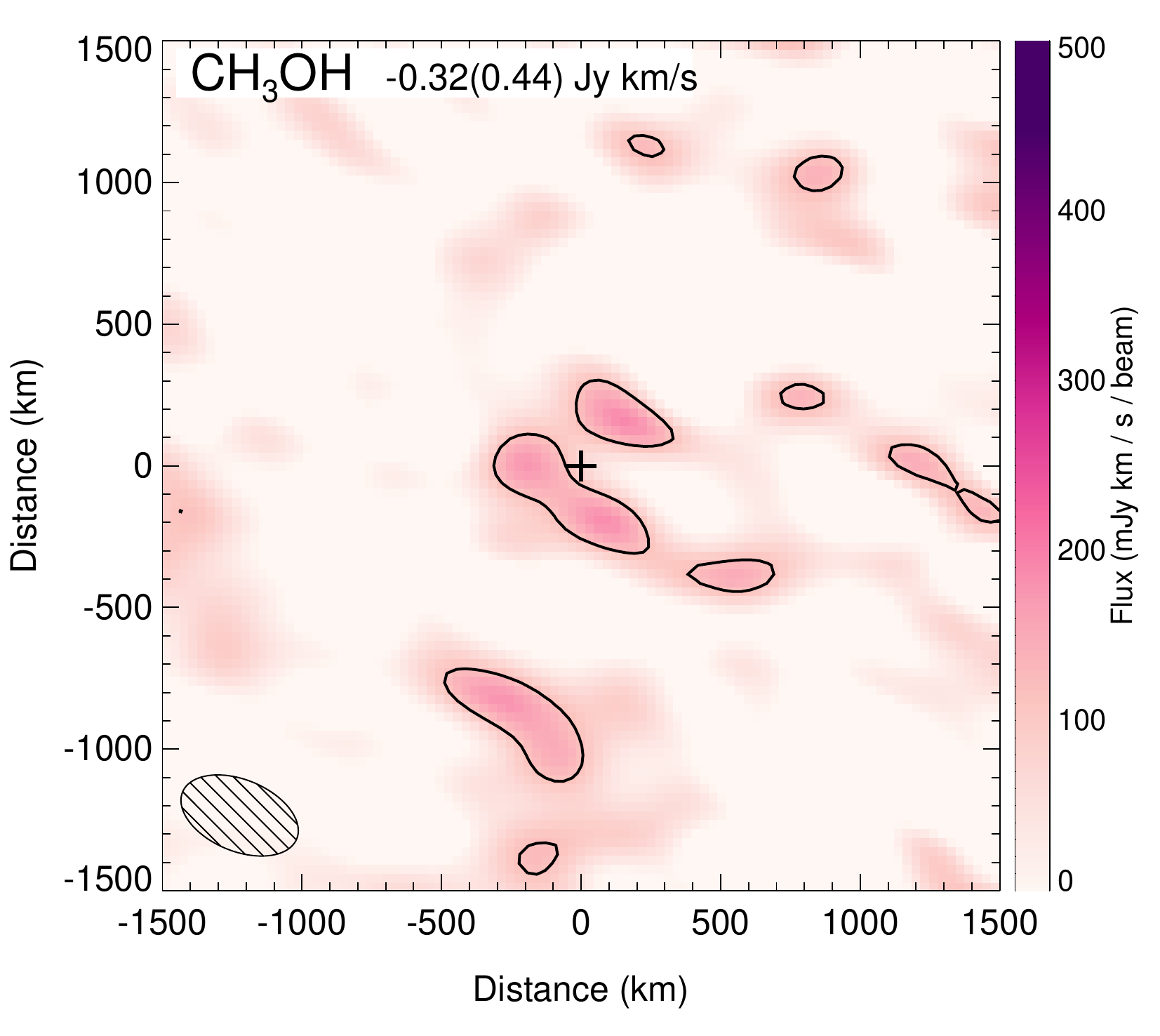}
\caption{Contour maps of the spectrally-integrated HNC (top row) and CH$_3$OH (bottom row) line fluxes observed in comet S1/ISON on 2013 November 16. Three time intervals are shown, each separated by 18 minutes. Contour intervals are $1.5\sigma$ for HNC and $2\sigma$ for CH$_3$OH. Integrated flux within a circular aperture of radius 1600~km centered on the comet is given at the top left of each map, { with $1\sigma$ error in parentheses}. Central crosses mark the same reference position relative to the comet in each map. Hatched ellipse indicates the spatial resolution. Axes are aligned with the equatorial system (celestial north is up). \label{fig:nov16_hnc}}
\end{figure*}

\begin{figure*}
\centering
UT 2013-11-16\ \ 10:18
\hspace{3.2cm}
UT 2013-11-16\ \ 10:35
\hspace{3.2cm}
UT 2013-11-16\ \ 10:53
\includegraphics[width=0.32\textwidth]{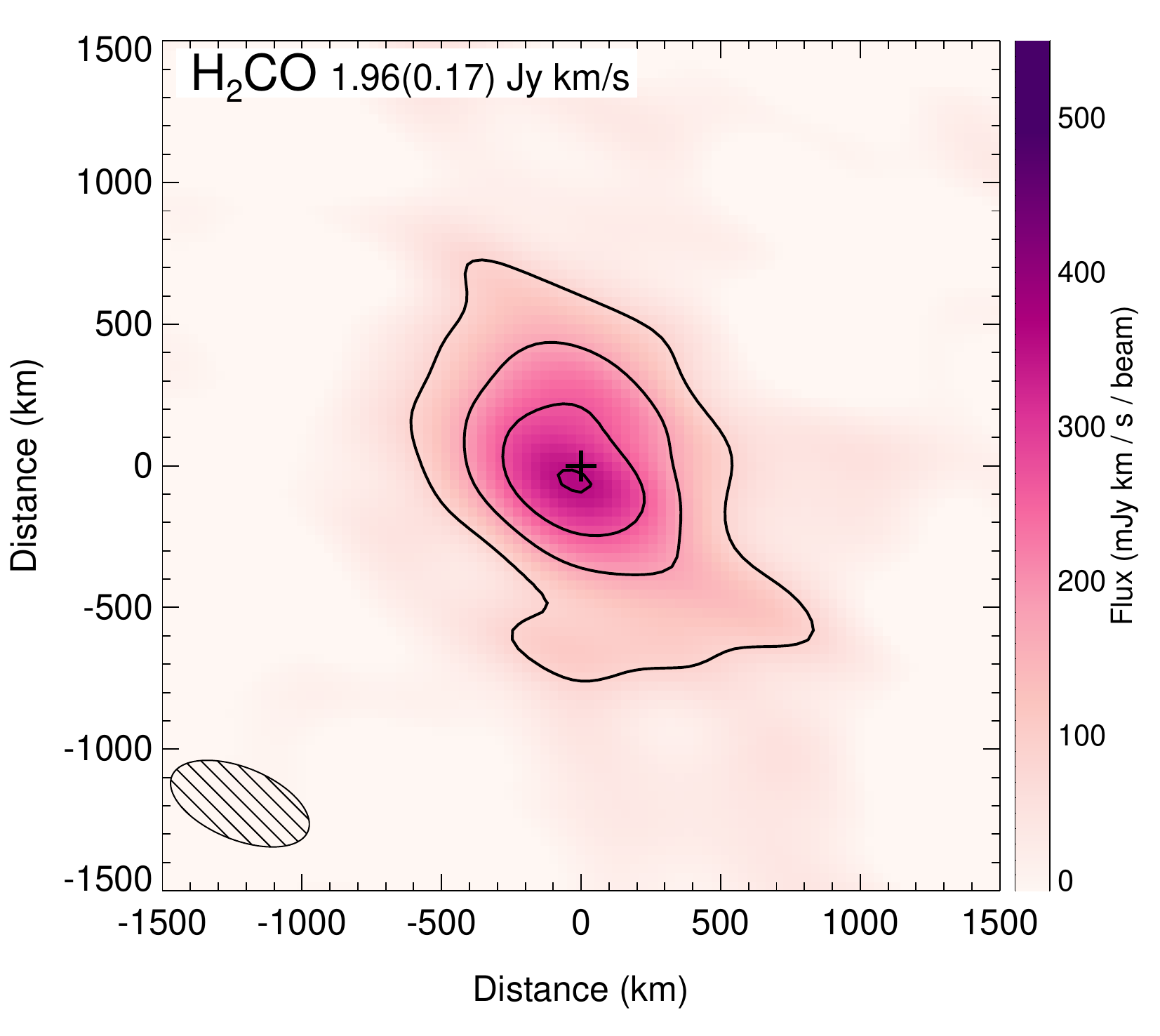}
\includegraphics[width=0.32\textwidth]{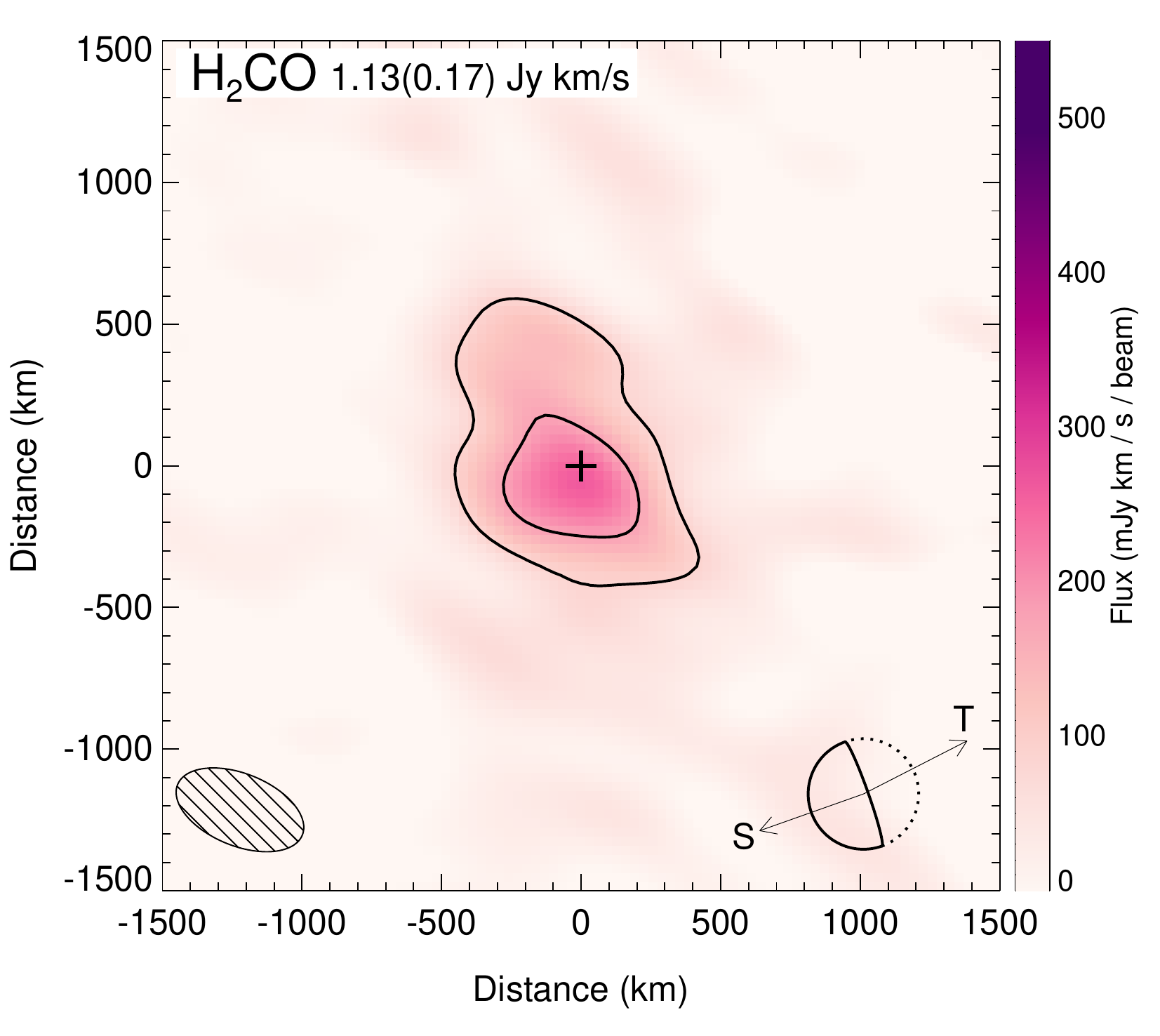}
\includegraphics[width=0.32\textwidth]{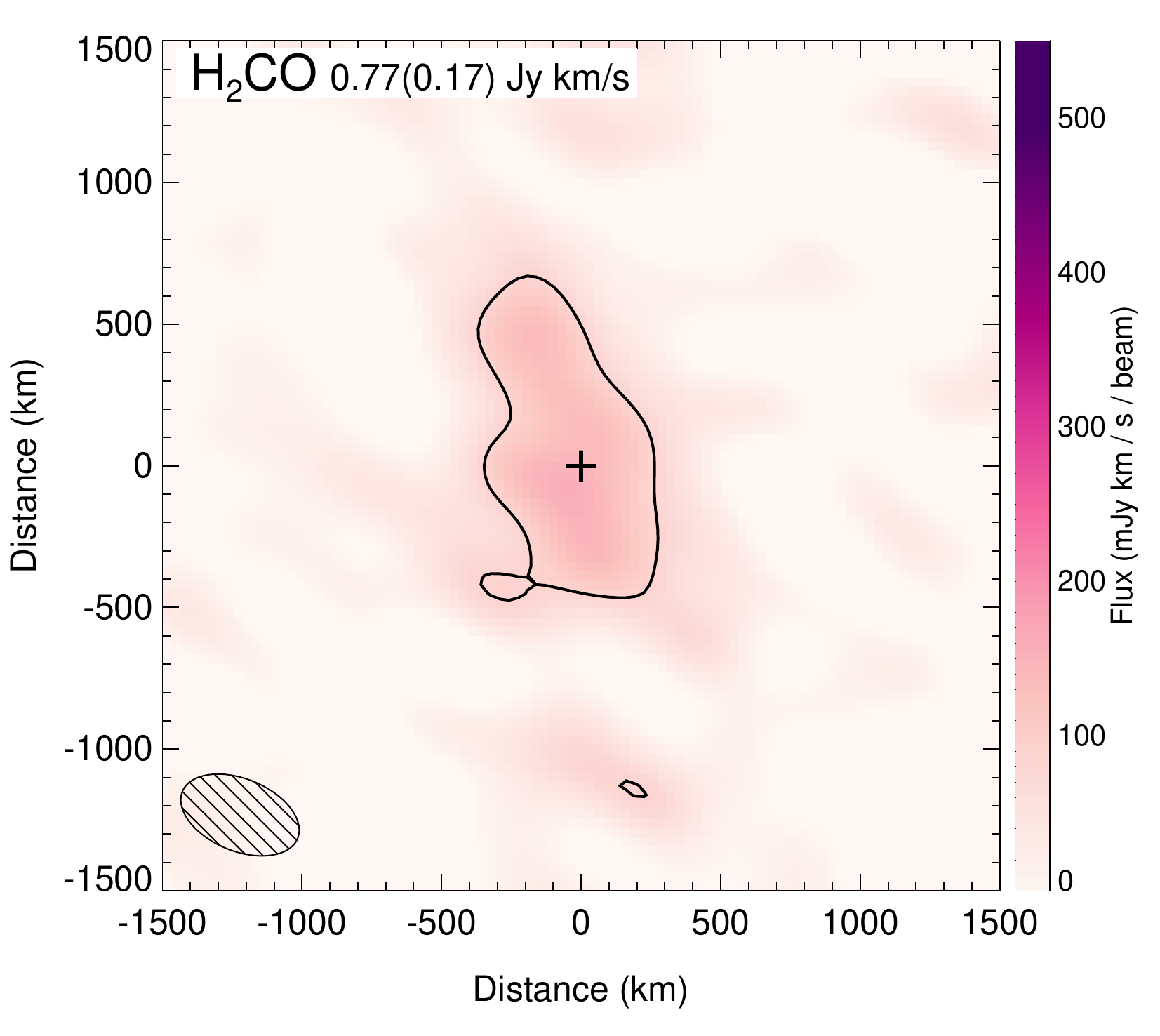}\\
\includegraphics[width=0.32\textwidth]{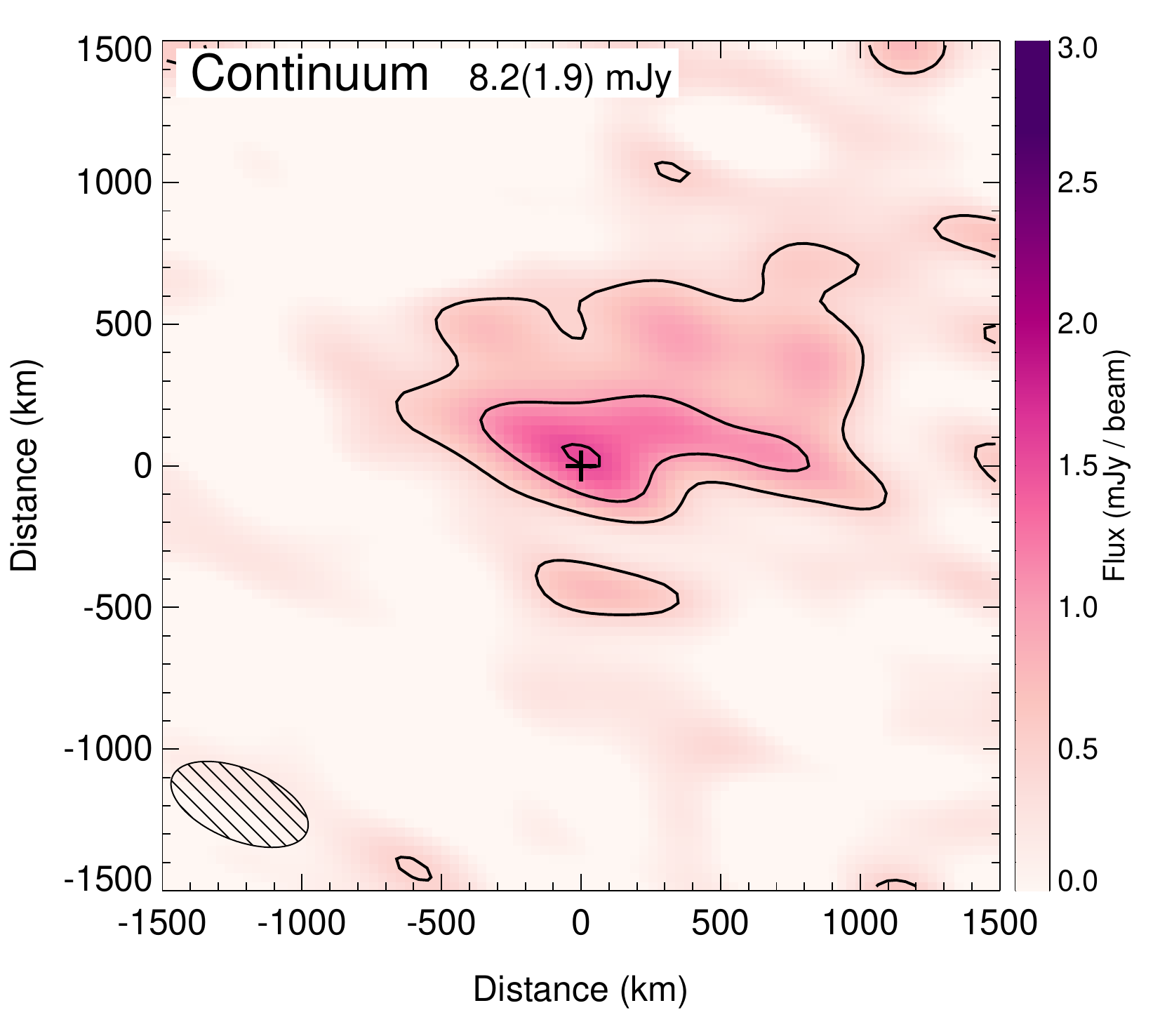}
\includegraphics[width=0.32\textwidth]{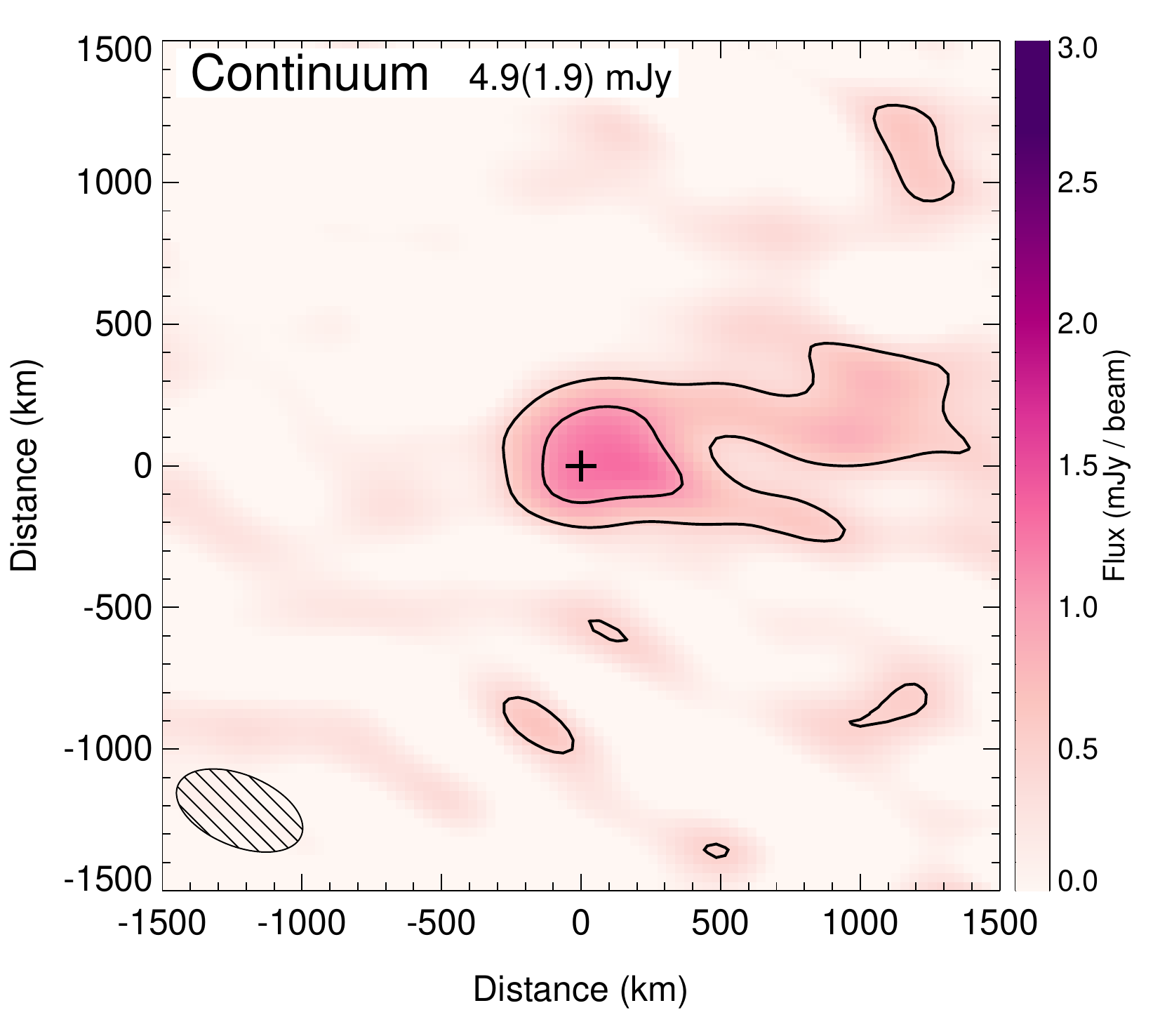}
\includegraphics[width=0.32\textwidth]{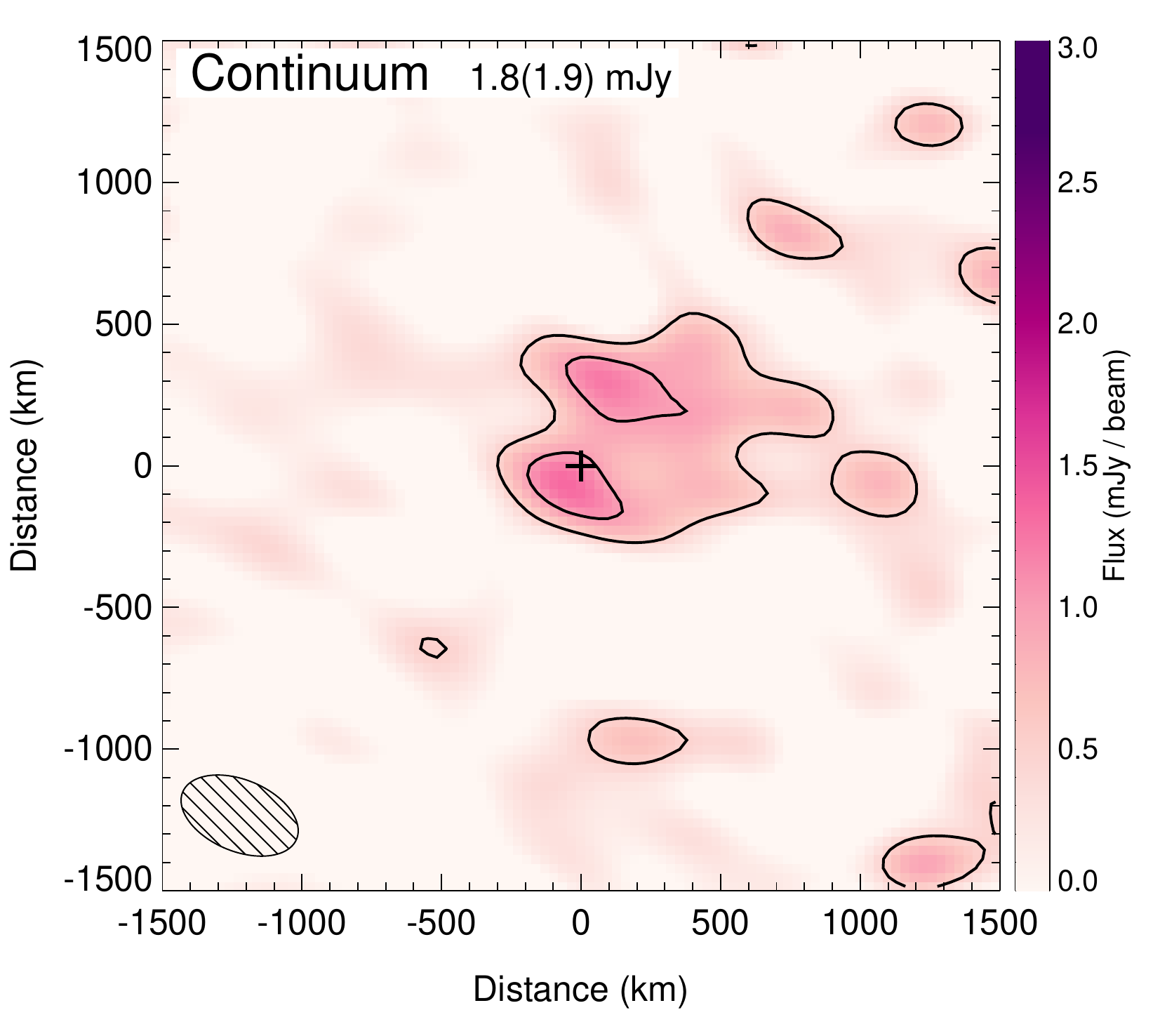}
\caption{As Fig. \ref{fig:nov16_hnc} but for H$_2$CO and continuum emission. Contour intervals are $4\sigma$ for H$_2$CO and $1\sigma$ for continuum. The comet's illumination phase and sky-projected vectors in the direction of the Sun and dust trail are shown bottom-right in the upper middle panel. \label{fig:nov16_h2co}}
\end{figure*}

\begin{figure*}
\centering
UT 2013-11-17\ \ 12:37
\hspace{3.2cm}
UT 2013-11-17\ \ 12:55
\hspace{3.2cm}
UT 2013-11-17\ \ 13:12
\includegraphics[width=0.32\textwidth]{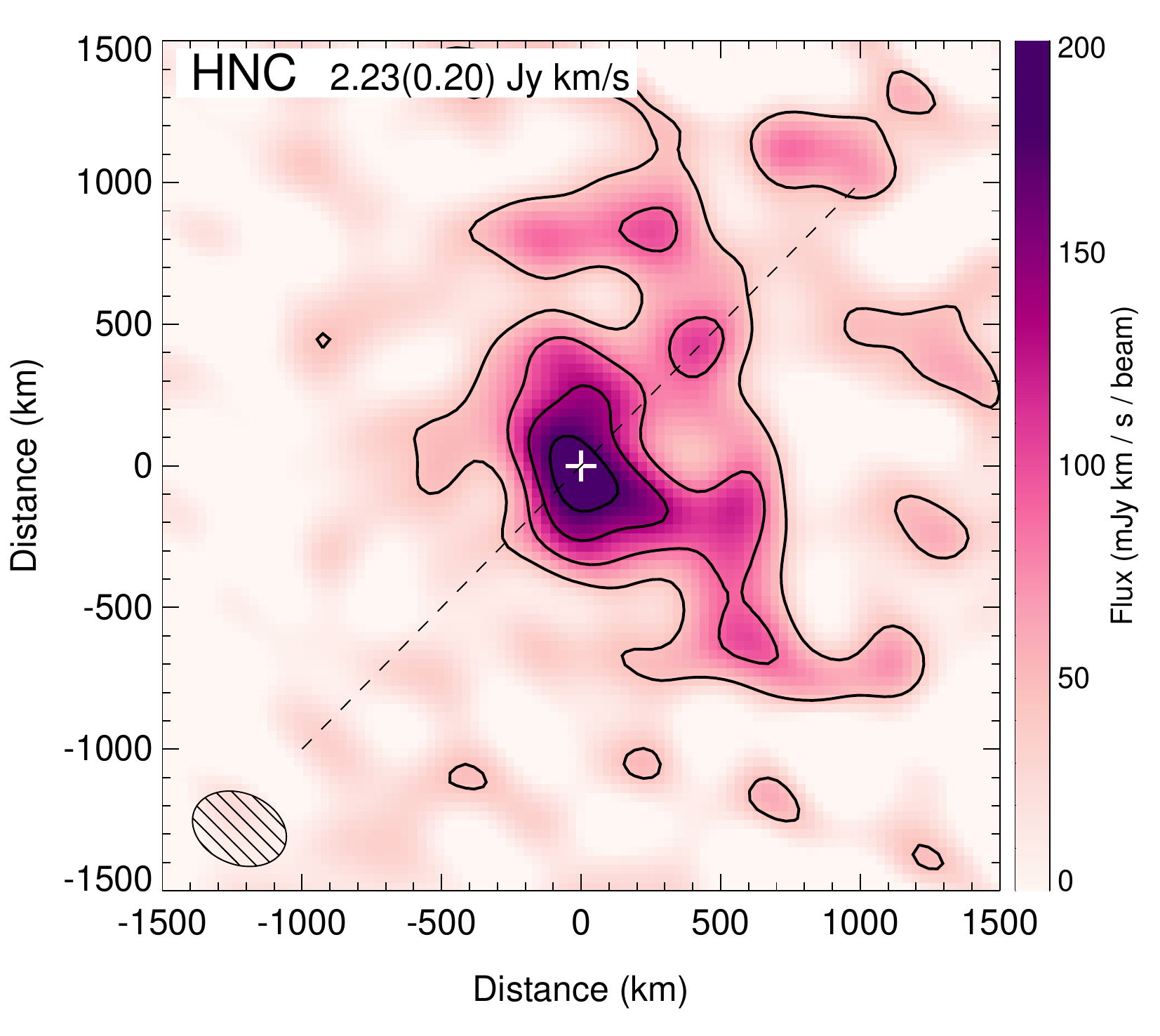}
\includegraphics[width=0.32\textwidth]{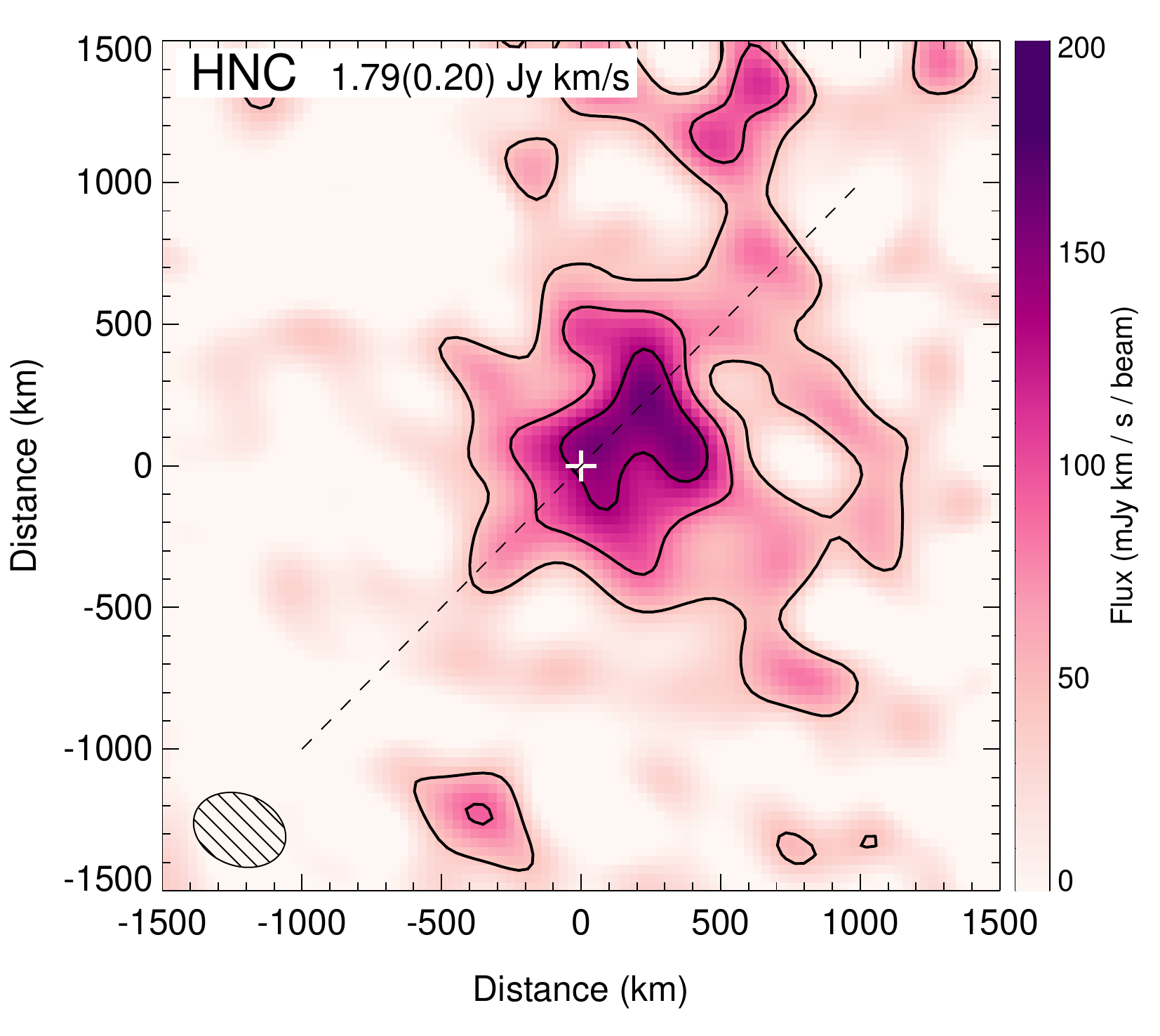}
\includegraphics[width=0.32\textwidth]{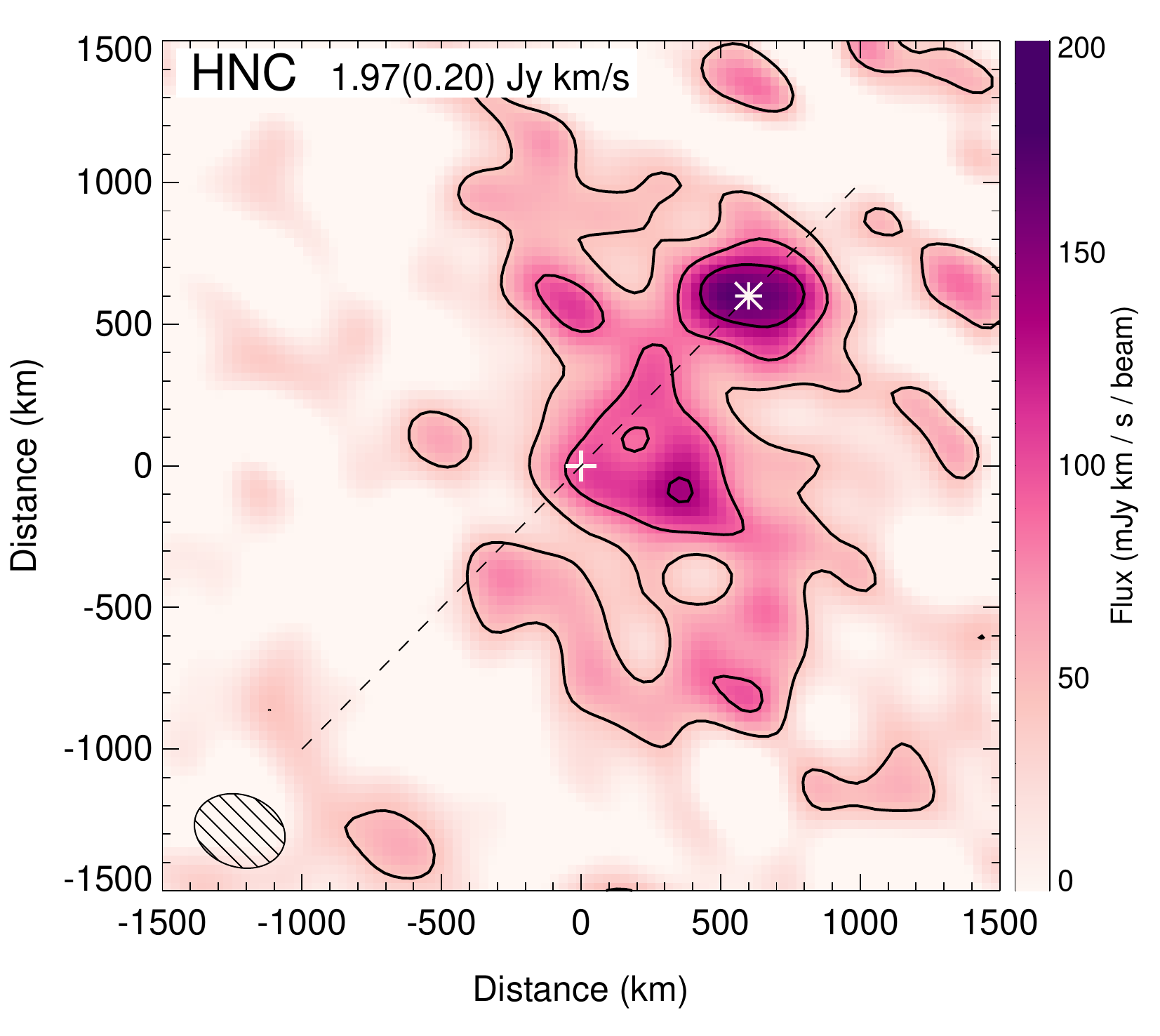}\\
\includegraphics[width=0.32\textwidth]{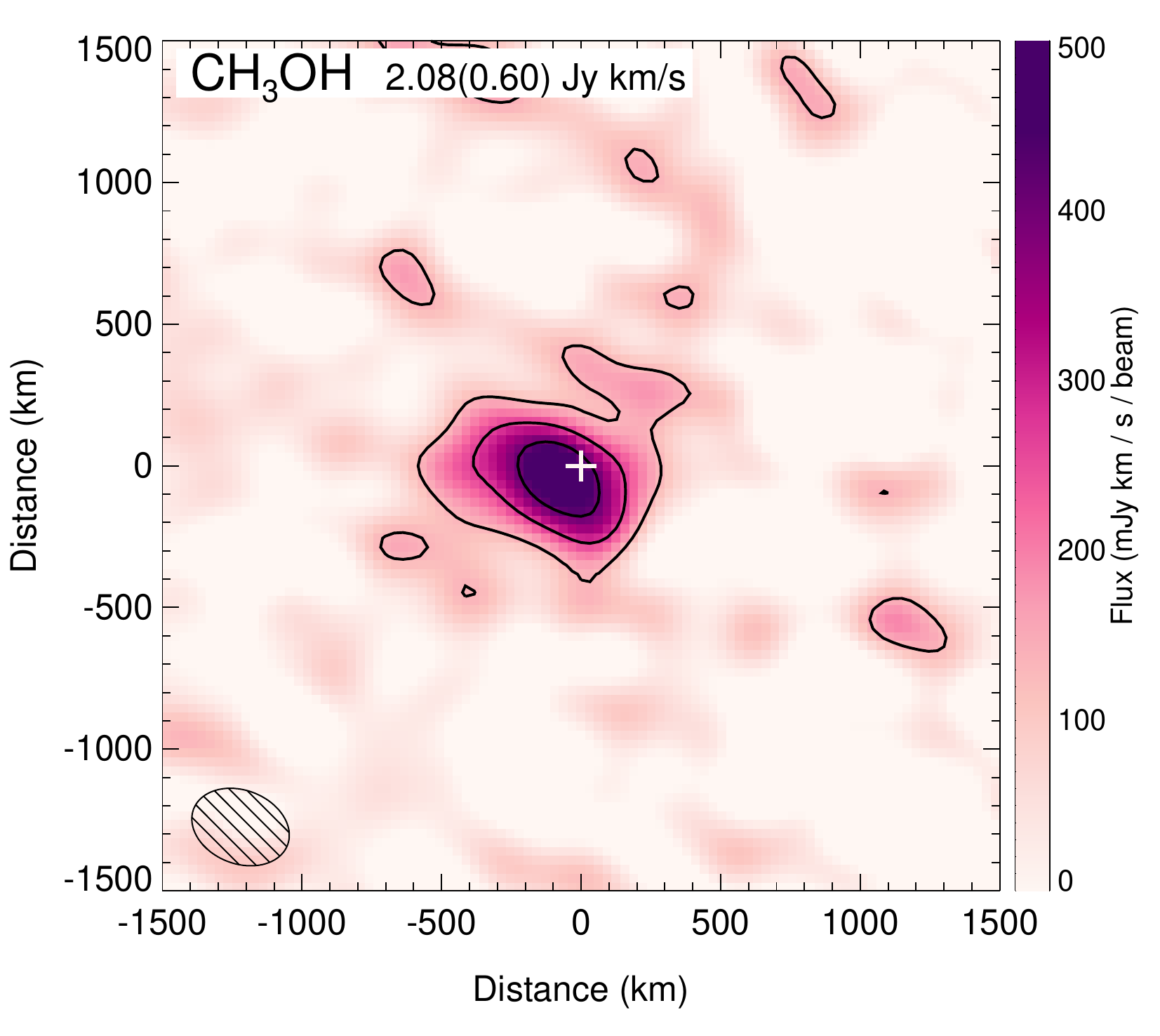}
\includegraphics[width=0.32\textwidth]{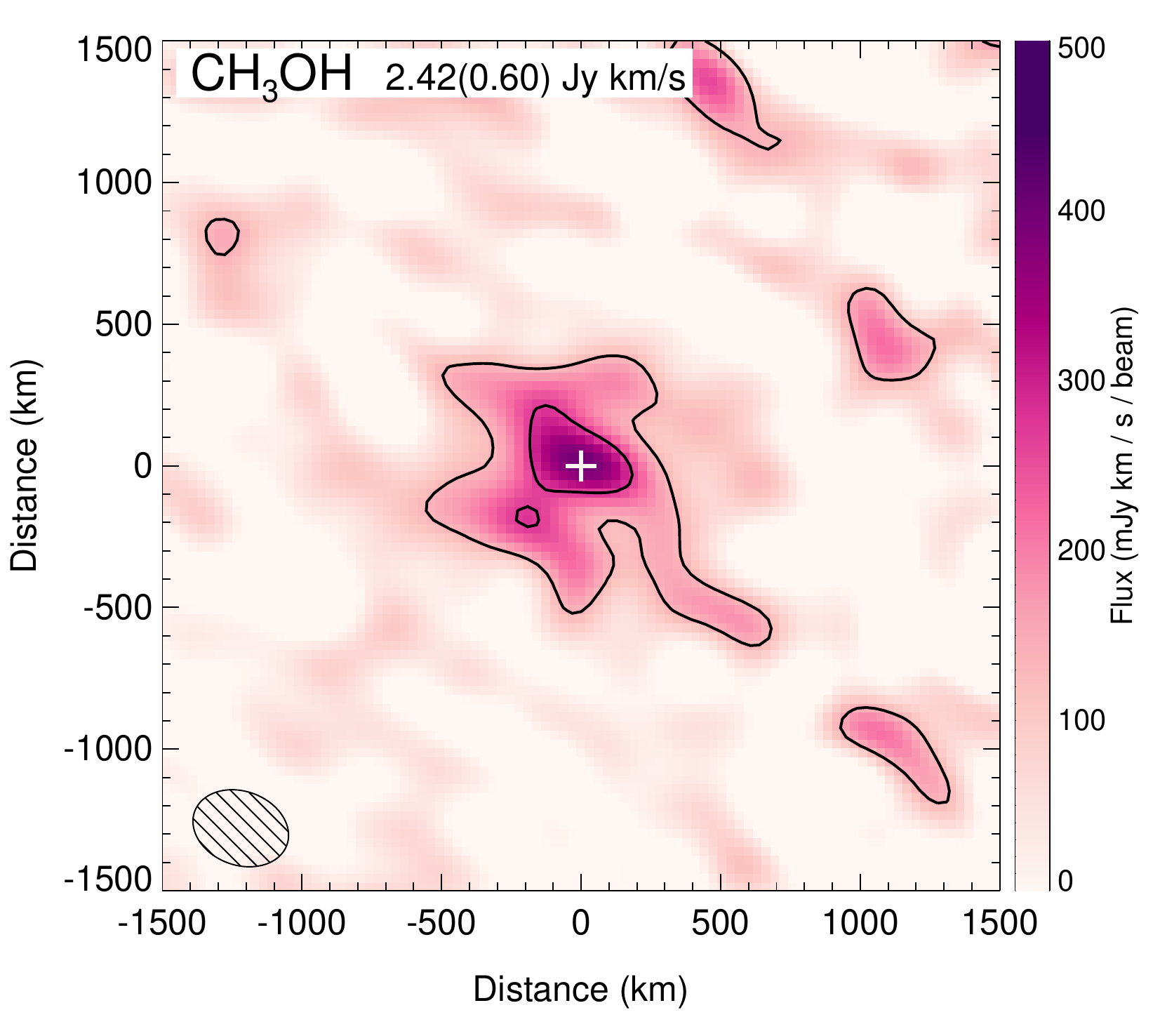}
\includegraphics[width=0.32\textwidth]{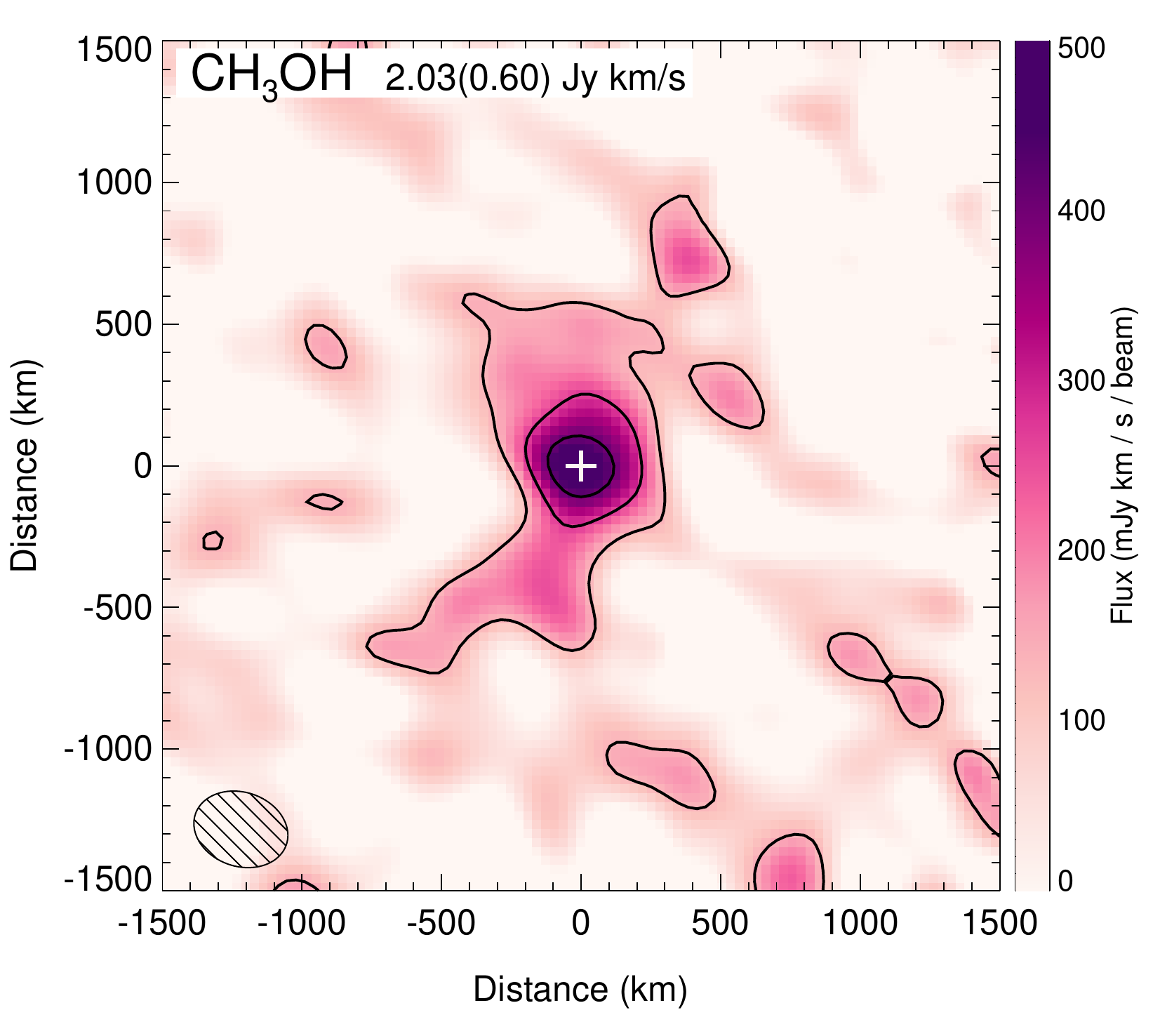}
\caption{As Fig. \ref{fig:nov16_hnc} but for the following day (2013 November 17). Asterisk in upper right panel indicates the position of an HNC clump and the dashed line is its putative outflow axis. Contour intervals are $1.5\sigma$ for HNC and $2\sigma$ for CH$_3$OH, where the RMS noise levels are $\sigma({\rm HNC})=30$~mJy\,\kms\ and $\sigma({\rm CH_3OH})=68$~mJy\,\kms. \label{fig:nov17_hnc}}
\end{figure*}

\begin{figure*}
\centering
UT 2013-11-17\ \ 12:37
\hspace{3.2cm}
UT 2013-11-17\ \ 12:55
\hspace{3.2cm}
UT 2013-11-17\ \ 13:12
\includegraphics[width=0.32\textwidth]{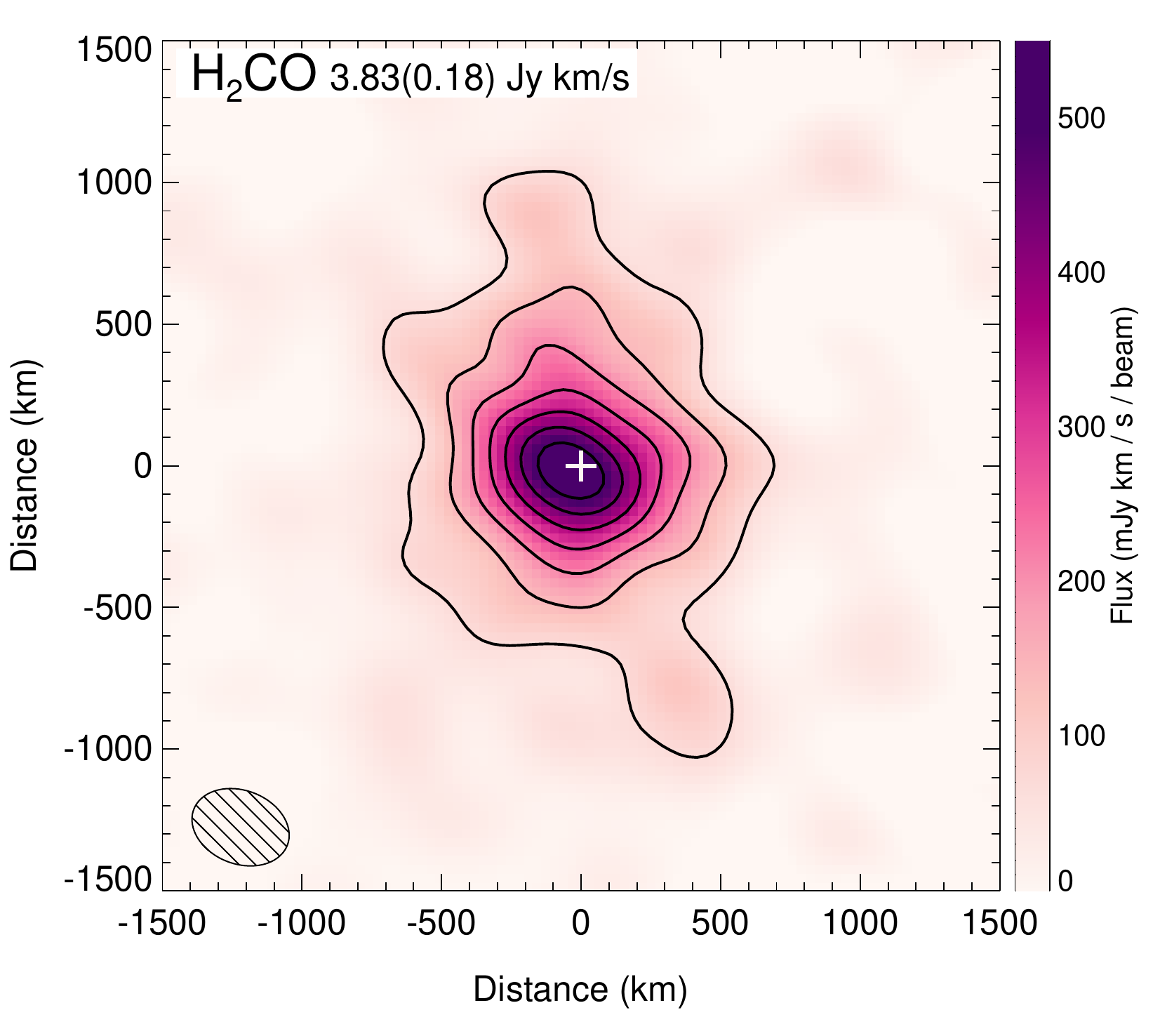}
\includegraphics[width=0.32\textwidth]{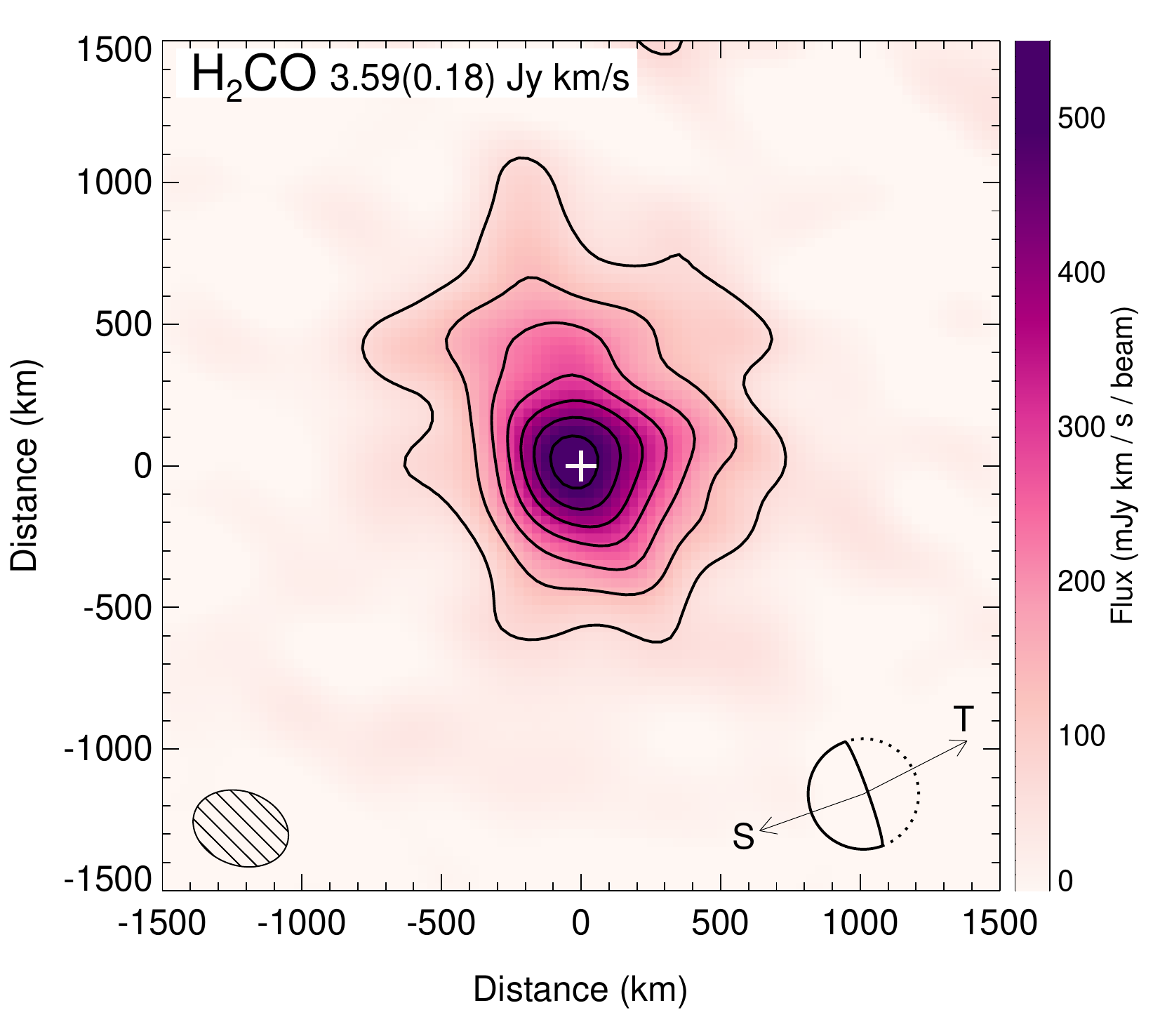}
\includegraphics[width=0.32\textwidth]{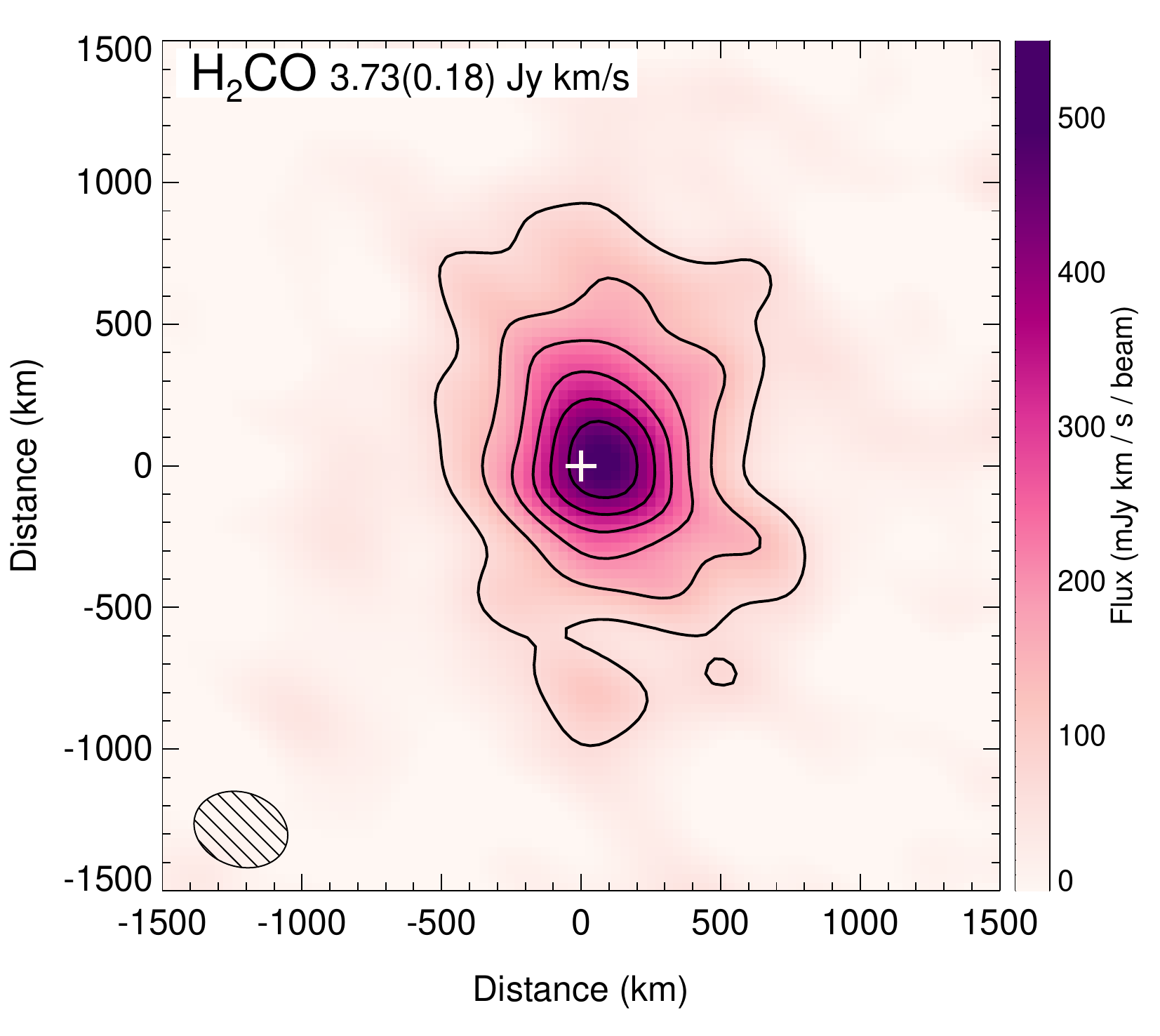}\\
\includegraphics[width=0.32\textwidth]{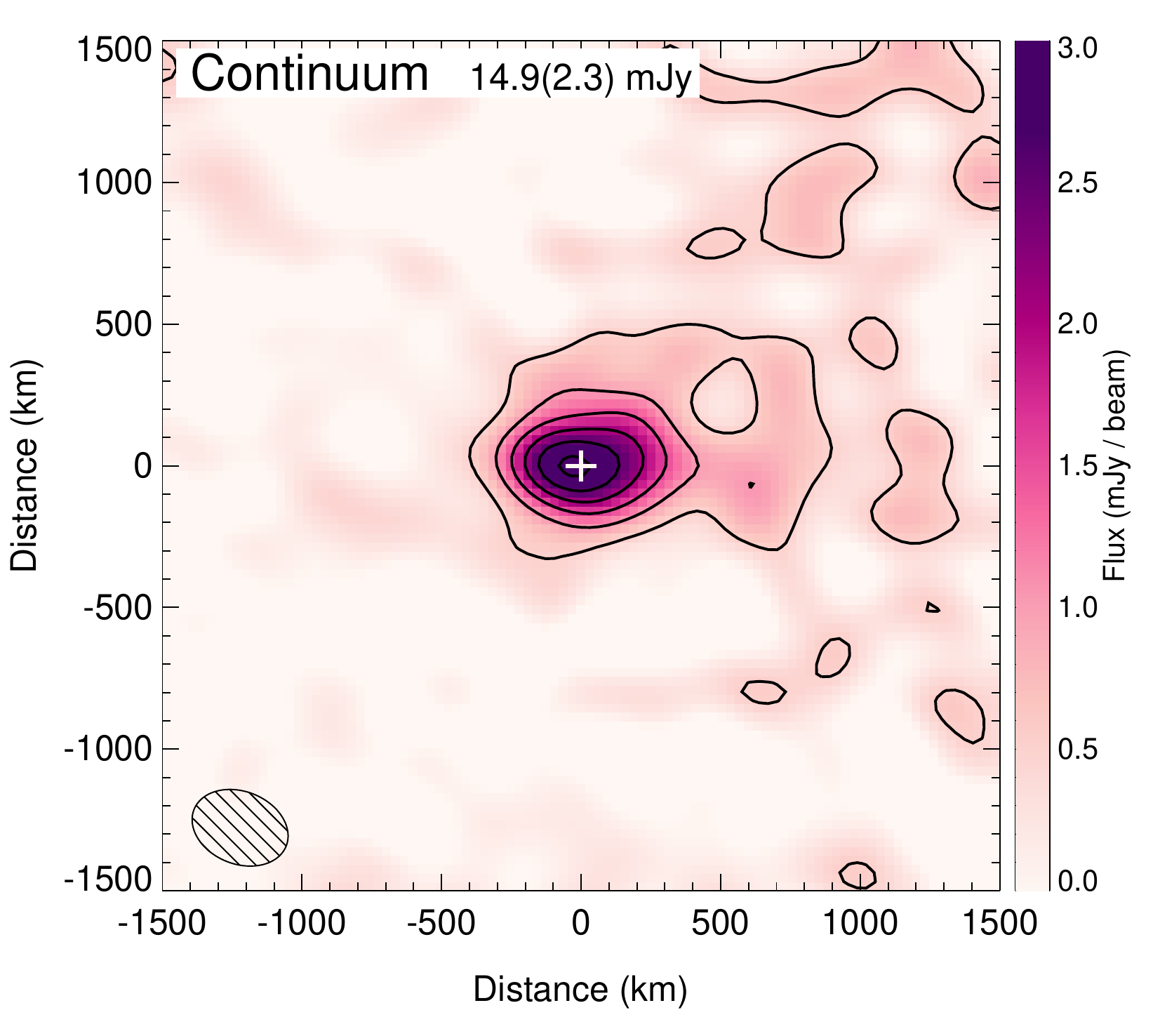}
\includegraphics[width=0.32\textwidth]{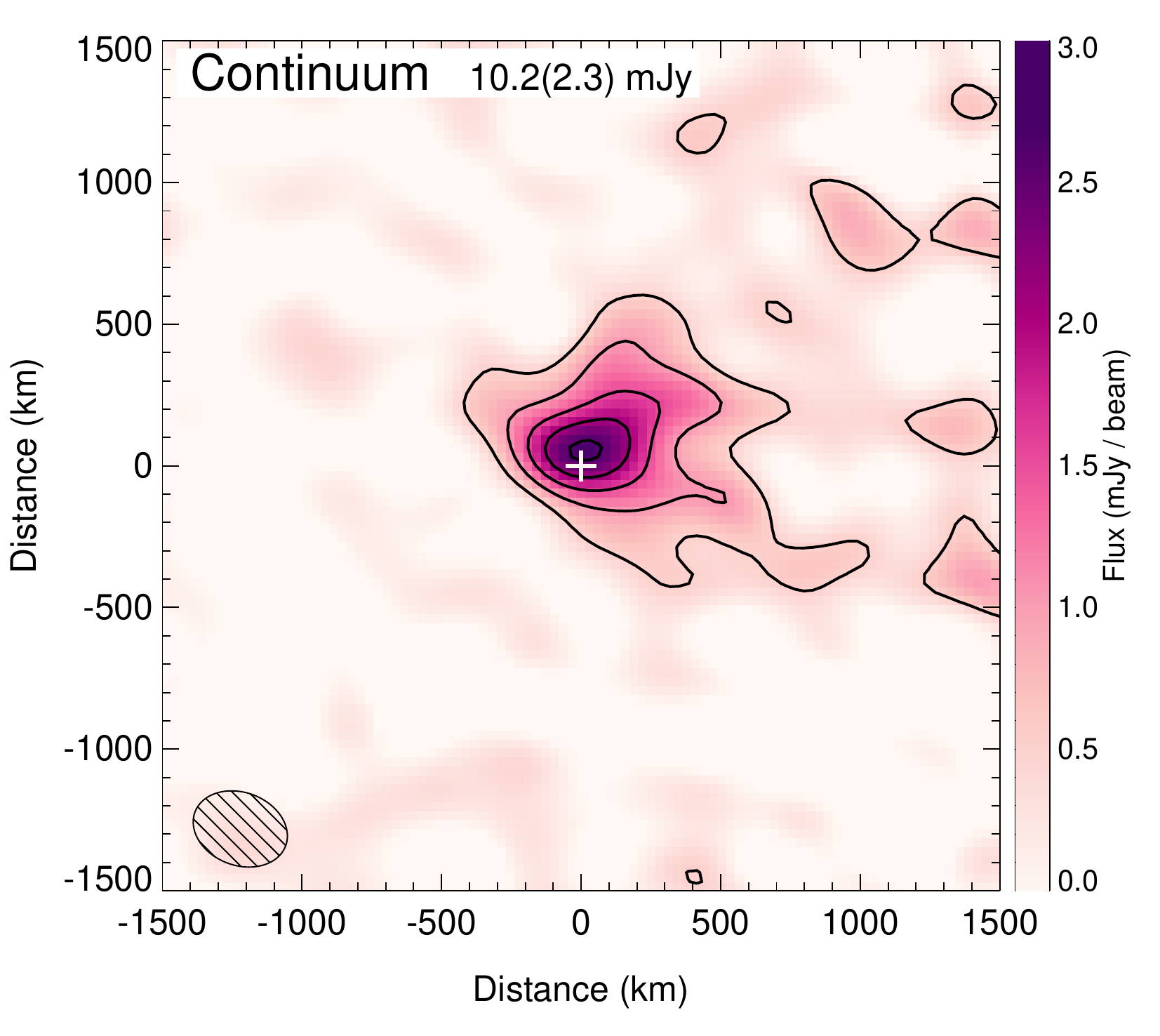}
\includegraphics[width=0.32\textwidth]{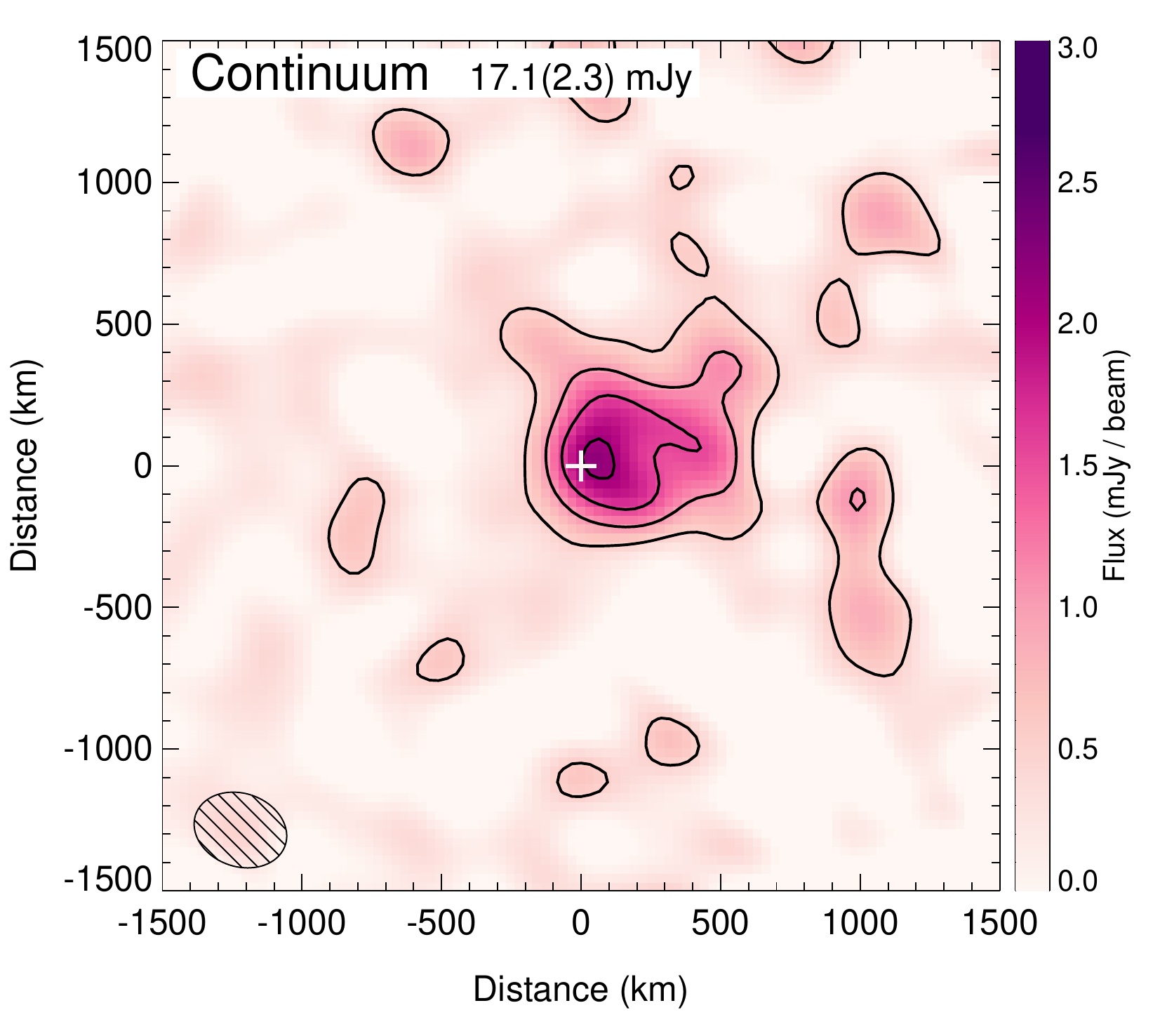}
\caption{As Fig. \ref{fig:nov16_h2co} but for the following day (2013 November 17). Contour intervals are $4\sigma$ for H$_2$CO and $1\sigma$ for continuum, where $\sigma({\rm H_2CO})=18$~mJy\,\kms\ and $\sigma({\rm continuum})=0.26$~mJy\,\kms.\label{fig:nov17_h2co}}
\end{figure*}

The molecular line data cubes were spectrally integrated over the full extent of detectable emission (between $-2$ and $+2$~\kms\ of line center), to produce maps for each molecule in each time period on November 16 (Figs. \ref{fig:nov16_hnc} and \ref{fig:nov16_h2co}) and November 17 (Figs. \ref{fig:nov17_hnc} and \ref{fig:nov17_h2co}). For consistency, common integration ranges and contour levels were applied across all time periods for a given species. For CH$_3$OH, several emission lines were detected (the $J_K=1_1-0_0$, $4_0-3_{-1}$, $7_2-6_1$ and $16_1-16_0$ transitions), and these data were summed together to produce a single image for each period, with improved signal-to-noise ratio.

For each species (plus continuum), the total flux was obtained by integrating over the emission maps within a circle of radius $2.5''$ (1600~km) centered on the origin. These fluxes are given in Table \ref{tab:lines}. Uncertainties (in parentheses) are statistical $1\sigma$ errors, calculated from the RMS of emission-free regions of the observed images.

\begin{table*}
\centering
\caption{Integrated spectral line and continuum fluxes as a function of time  \label{tab:lines}}
\begin{tabular}{lccccrr}
\hline\hline
Species&Period$^a$&Transition&Frequency&E$_u$&\multicolumn{2}{c}{Flux$^b$}\\
\cline{6-7}
&&&(GHz)&(K)&Nov. 16&Nov. 17\\
\hline
HNC         &$1$   &$4-3$&362.630&43.5&$1.03\pm20$&$2.23\pm20$\\
HNC         &$2$&$4-3$&362.630&43.5&$0.21\pm20$&$1.79\pm20$\\
HNC         &$3$&$4-3$&362.630&43.5&$0.17\pm20$&$1.97\pm20$\\[1mm]
CH$_3$OH    &$1$   &multi$^c$&350-364&17-333&$1.02\pm44$&$2.08\pm60$\\
CH$_3$OH    &$2$&multi$^c$&350-364&17-333&$0.21\pm44$&$2.42\pm60$\\
CH$_3$OH    &$3$&multi$^c$&350-364&17-333&$-0.32\pm44$&$2.03\pm60$\\[1mm]
H$_2$CO     &$1$ &$5_{1,5}-4_{1,4}$&351.769&62.5&$1.96\pm17$&$3.83\pm18$\\
H$_2$CO     &$2$&$5_{1,5}-4_{1,4}$&351.769&62.5&$1.13\pm17$&$3.59\pm18$\\
H$_2$CO     &$3$&$5_{1,5}-4_{1,4}$&351.769&62.5&$0.77\pm17$&$3.73\pm18$\\[1mm]
Continuum   &$1$ &\dots&357&\dots&$8.2\pm1.9$&$14.9\pm2.3$\\
Continuum   &$2$&\dots&357&\dots&$4.9\pm1.9$&$10.2\pm2.3$\\
Continuum   &$3$&\dots&357&\dots&$1.8\pm1.9$&$17.1\pm2.3$\\
\hline
\end{tabular}
\\
\parbox{1.1\columnwidth}{\footnotesize 
\vspace*{1mm}
$^a$ Time period --- see section \ref{sec:obs}.\\
$^b$ Integrated line flux within a 2.5$''$-radius circular aperture centered on the comet; $1\sigma$ errors on trailing digits given in parentheses. Units are Jy \kms\ for the molecular lines and mJy for continuum.\\
$^c$ The CH$_3$OH maps are a sum over four transitions: $4_0-3_{-1}$ (350687.7~MHz), $1_1-0_0$ (350905.1~MHz), $16_1-16_0$ (363440.4~MHz) and $7_2-6_1$ (363739.9~MHz).
\\

}
\end{table*}

\subsection{2013 November 16}

Between the start and finish of our observations on November 16, a dramatic drop in total integrated flux was observed in comet ISON for all spectral lines and the continuum. A calibration error or problem with the telescope array can be ruled out as the source of this variation because our images of the phase calibrator 3C\,279 (less than 10$^{\circ}$ away on the sky) showed a uniform intensity point-source over throughout the observations. As stated in Section \ref{sec:obs}, the antenna phase fluctuations were small enough to be easily tracked and corrected between visits to the phase calibrator; resulting in a stable flux response as a function of time.

The HNC map for period 1, observed on Nov. 16 (starting at UT 2013-11-16 10:10; see Fig. \ref{fig:nov16_hnc}) shows a peak near the nominal nucleus position (indicated with a white cross), and a second, elongated structure extending away towards the north, which appears to be partially separated from the main peak. This northerly elongated clump is reminiscent of the HNC streams/jets identified by \citet{cor14} on Nov. 17, indicating that these peculiar structures were not a temporally isolated phenomenon unique to those observations. The HNC flux on Nov. 16 fell from $0.7\pm0.2$~Jy\,\kms\ to $0.1\pm0.2$~Jy\,\kms\ during the 35 minutes between periods 1 and 3. The corresponding reduction in signal-to-noise ratio precludes the possibility of tracking the time-evolution of the HNC clumpy structure on this date.

A significant reduction in flux as a function of time was also observed for CH$_3$OH, but again, these maps suffer from low signal-to-noise. The H$_2$CO maps contain more flux, showing a clear central peak and relatively symmetric spatial distribution consistent with isotropic outgassing. The H$_2$CO distribution remained centrally peaked while the flux fell in an approximately linear fashion by 52-72\% between periods 1-3.

\begin{figure}
\centering
UT 2013-11-16\\
\includegraphics[width=\columnwidth]{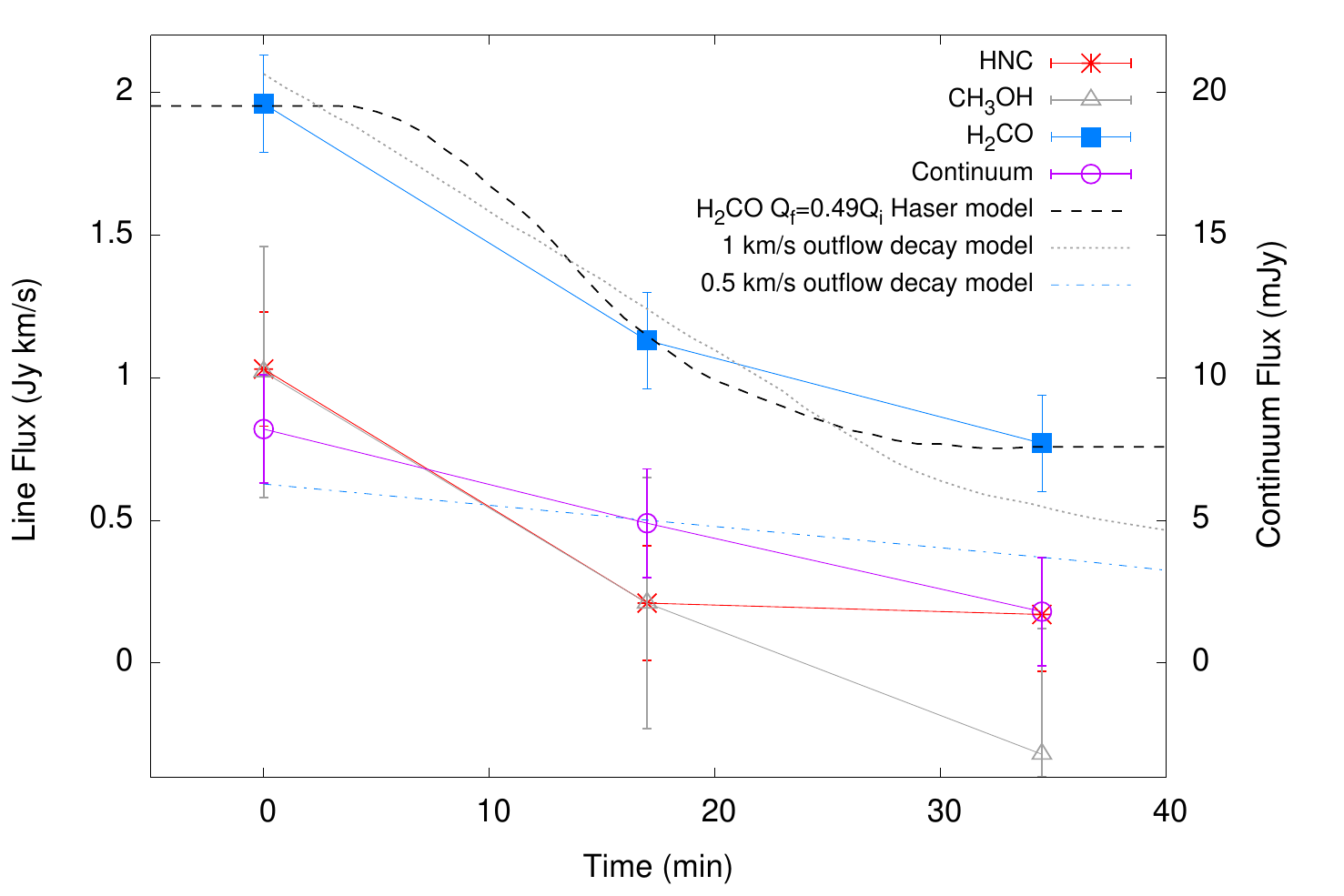}\\[3mm]
UT 2013-11-17\\
\includegraphics[width=\columnwidth]{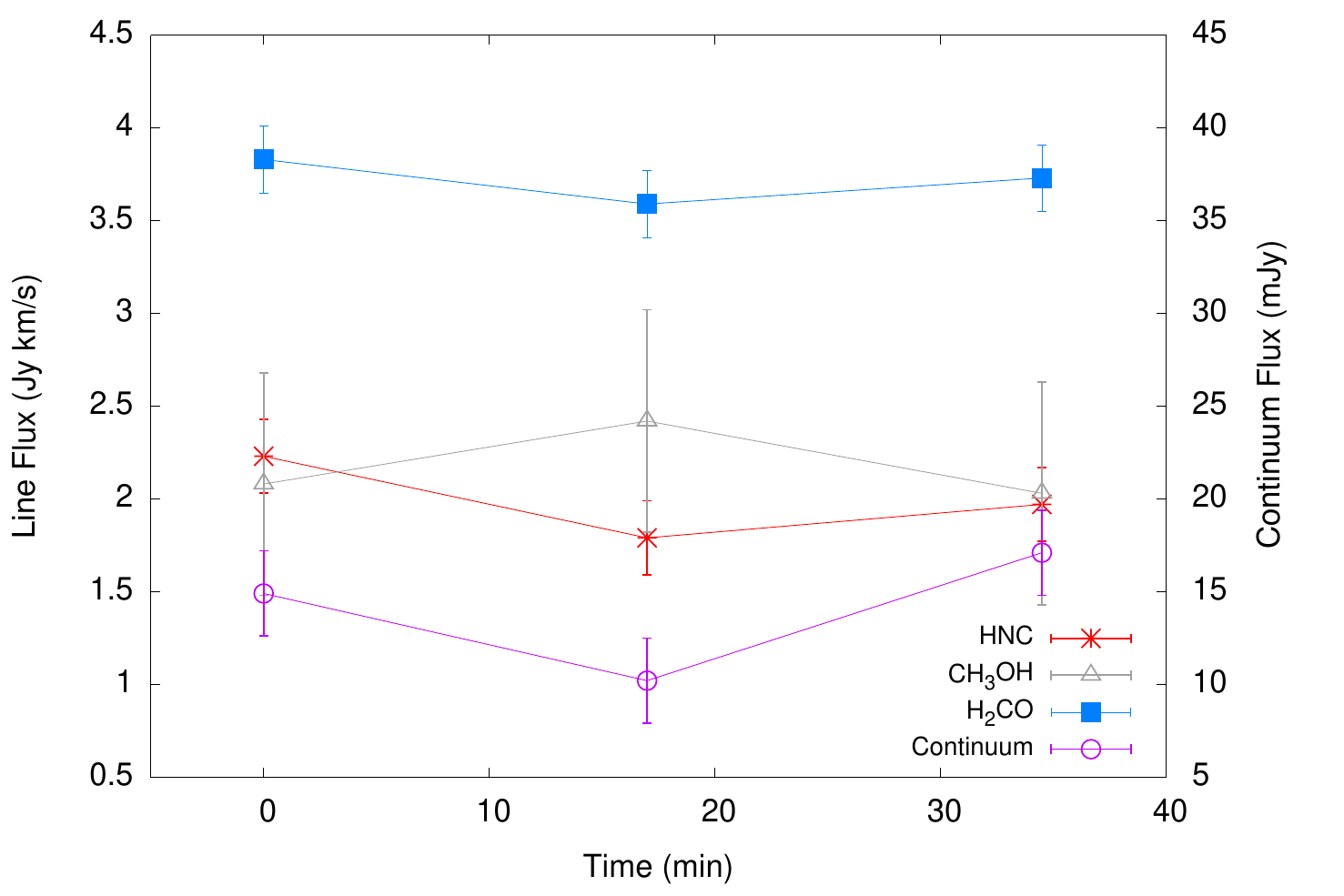}
\caption{Total line flux (left ordinate) and continuum flux (right ordinate) as a function of time after the start of observations on Nov 16 (top panel) and Nov 17 (bottom panel). The result of a simulated H$_2$CO outflow model (with gas production rate falling to $Q_f=0.49Qi$ after 3.5 minutes) is shown with a black dashed line. Calculated flux decay curves due to coma expansion following an instantaneous halt to cometary production are also shown for outflow velocities of 1~\kms\ and 0.5~\kms. \label{fig:fluxvstime}}
\end{figure}

In the 0.8~mm continuum maps (shown in the lower panels of Fig. \ref{fig:nov16_h2co}), the emission contours show a clear elongation towards the upper right of the figure, consistent with the approximate direction of the comet's orbital trail and the anti-sunward vector. Similar to the molecular species, the integrated continuum flux also exhibited an apparently linear decrease with time, falling by at least 42\% during our observations, and shows that the rapid decline in comet ISON's activity was not limited to only the gaseous component of the coma. Fig. \ref{fig:fluxvstime} (upper panel) shows a plot of the gas and continuum flux measurements as a function of time (including $1\sigma$ error bars), where the declining flux trends are \mbox{clearly
apparent}.

\begin{figure*}
\centering
\hspace{0.5cm}
{UT 2013-11-17\ \ 12:37}
\hspace{3.2cm}
{UT 2013-11-17\ \ 12:55}
\hspace{3.2cm}
{\footnotesize UT 2013-11-17\ \ 13:12}
\includegraphics[width=0.32\textwidth]{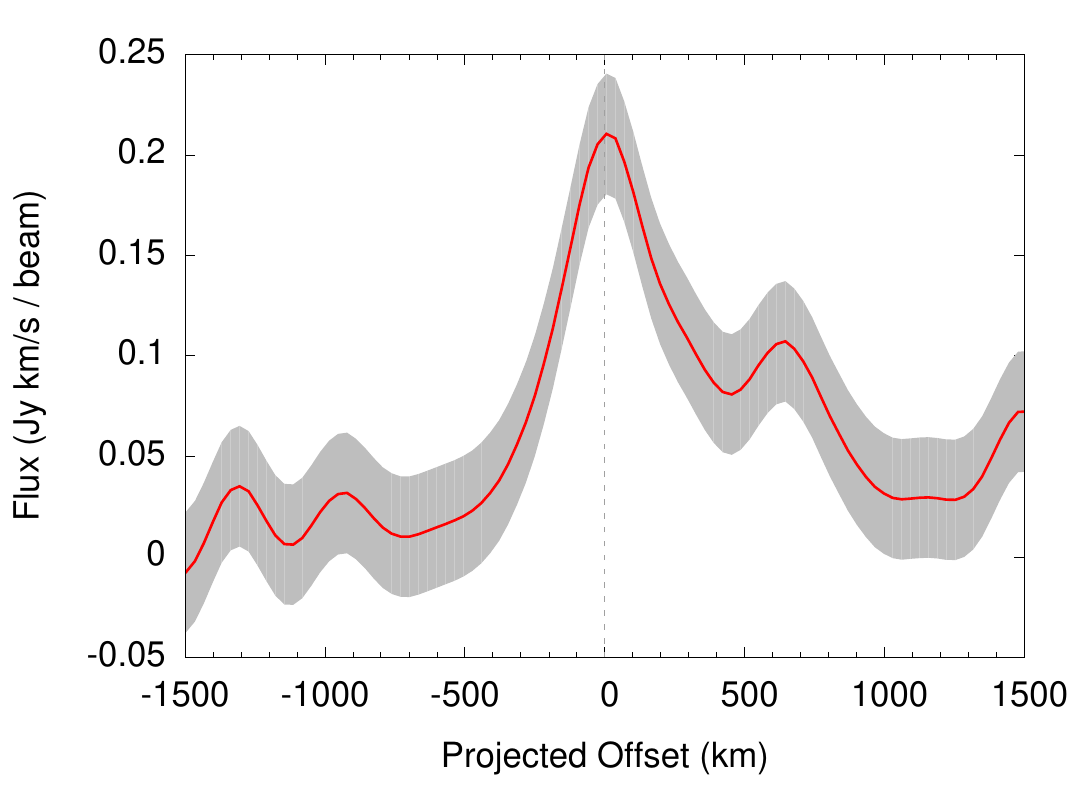}
\includegraphics[width=0.32\textwidth]{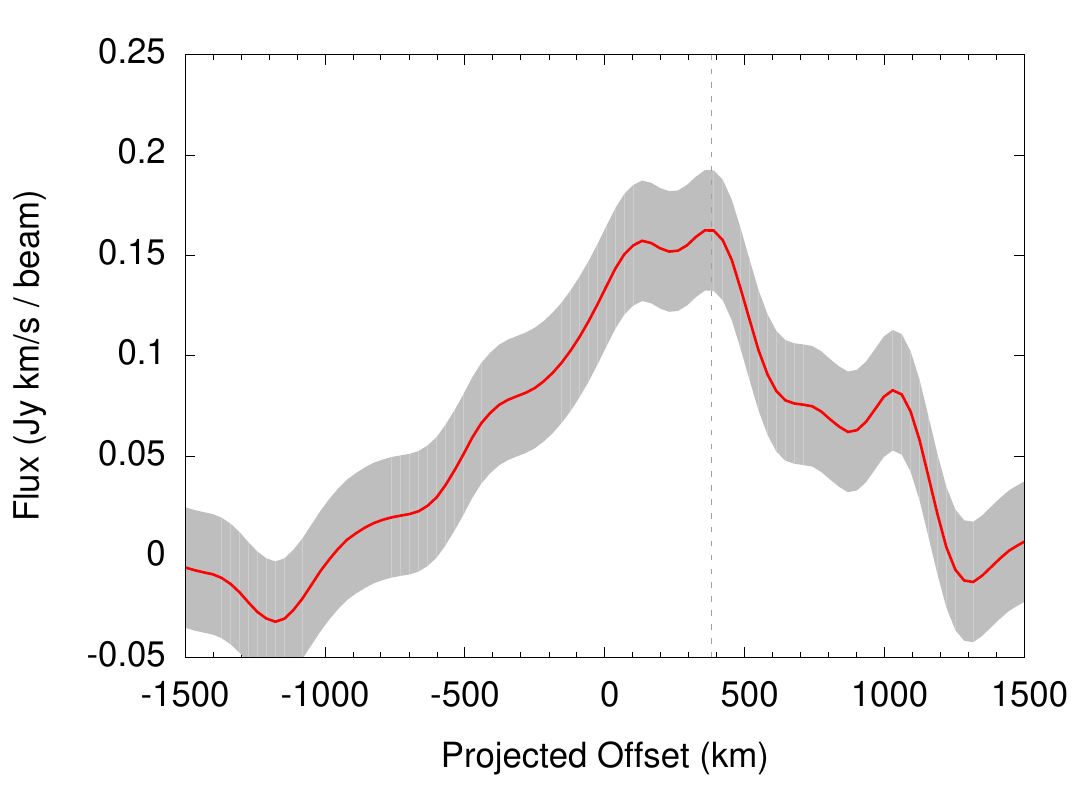}
\includegraphics[width=0.32\textwidth]{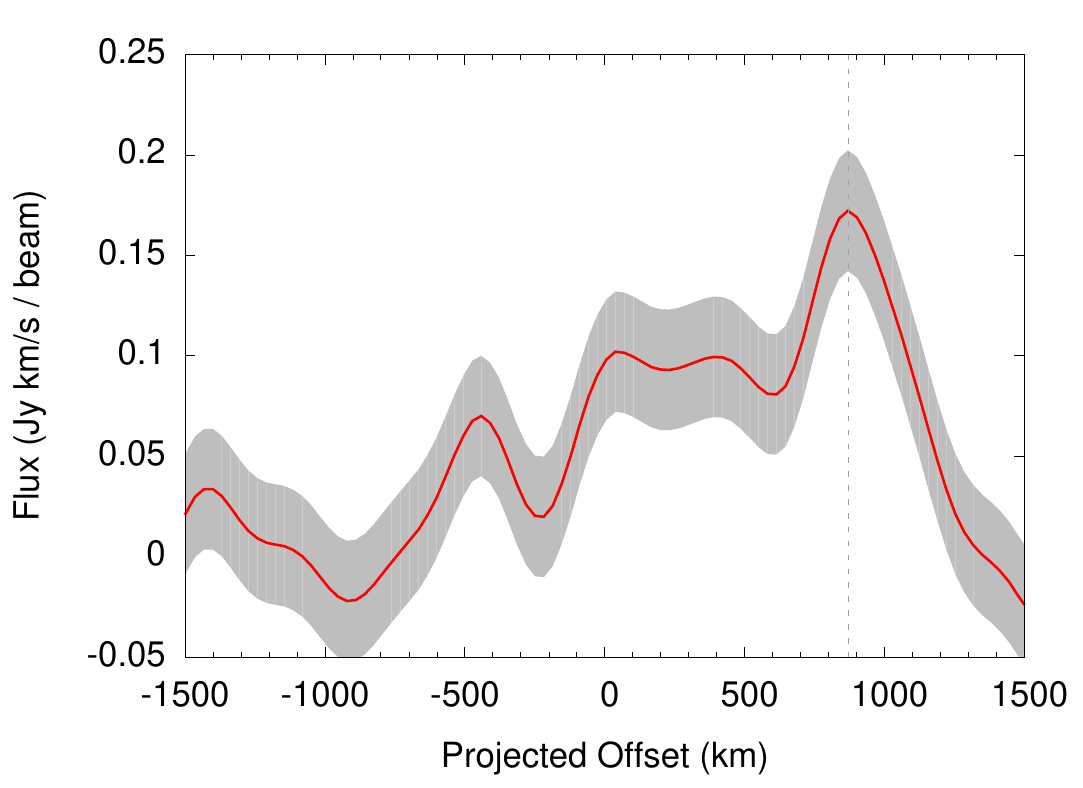}\\
\caption{Spatial cuts through the HNC maps at the top of Fig. \ref{fig:nov17_hnc}, showing the flux as a function of distance along the dashed black line (passing through the continuum peak and HNC clump marked '$\ast$'). Vertical dashed lines show the peak flux position for each period. \label{fig:profiles}}
\end{figure*}

\begin{figure*}
\centering
\hspace{0.6cm}
{UT 2013-11-17\ \ 12:37}
\hspace{2.8cm}
{UT 2013-11-17\ \ 12:55}
\hspace{2.8cm}
{\footnotesize UT 2013-11-17\ \ 13:12}\\[1mm]
\includegraphics[width=0.34\textwidth]{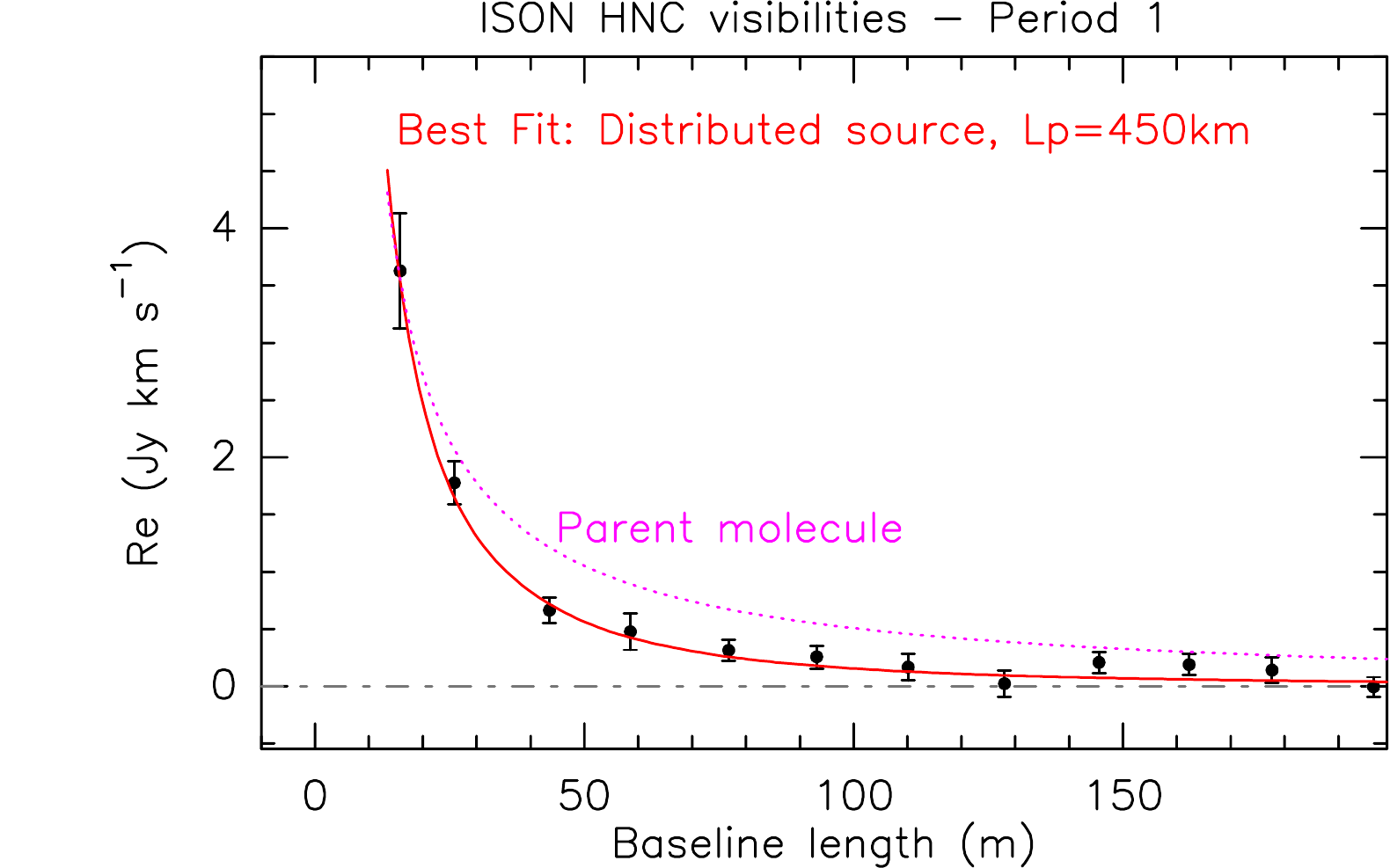}
\includegraphics[width=0.303\textwidth]{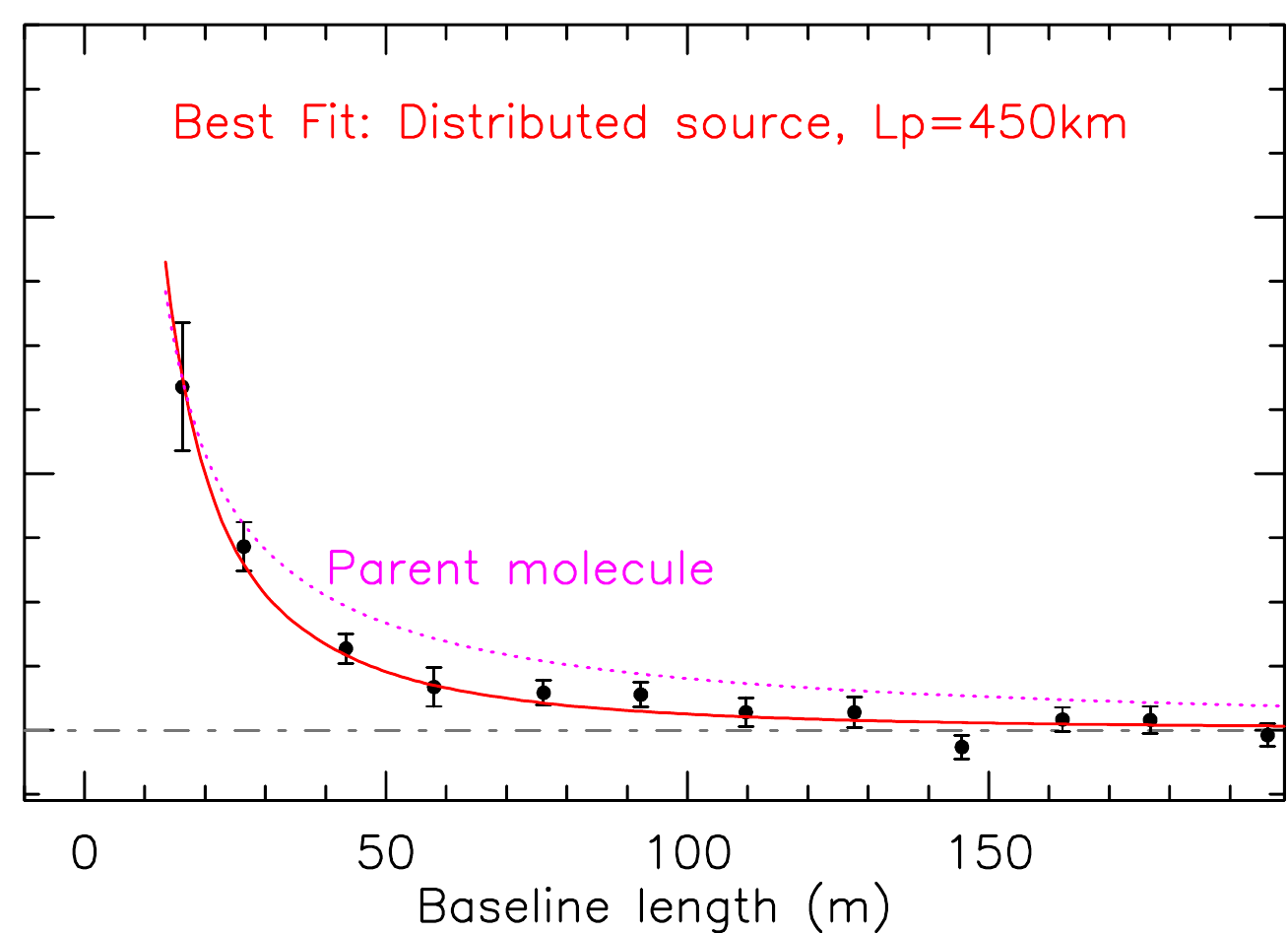}
\includegraphics[width=0.303\textwidth]{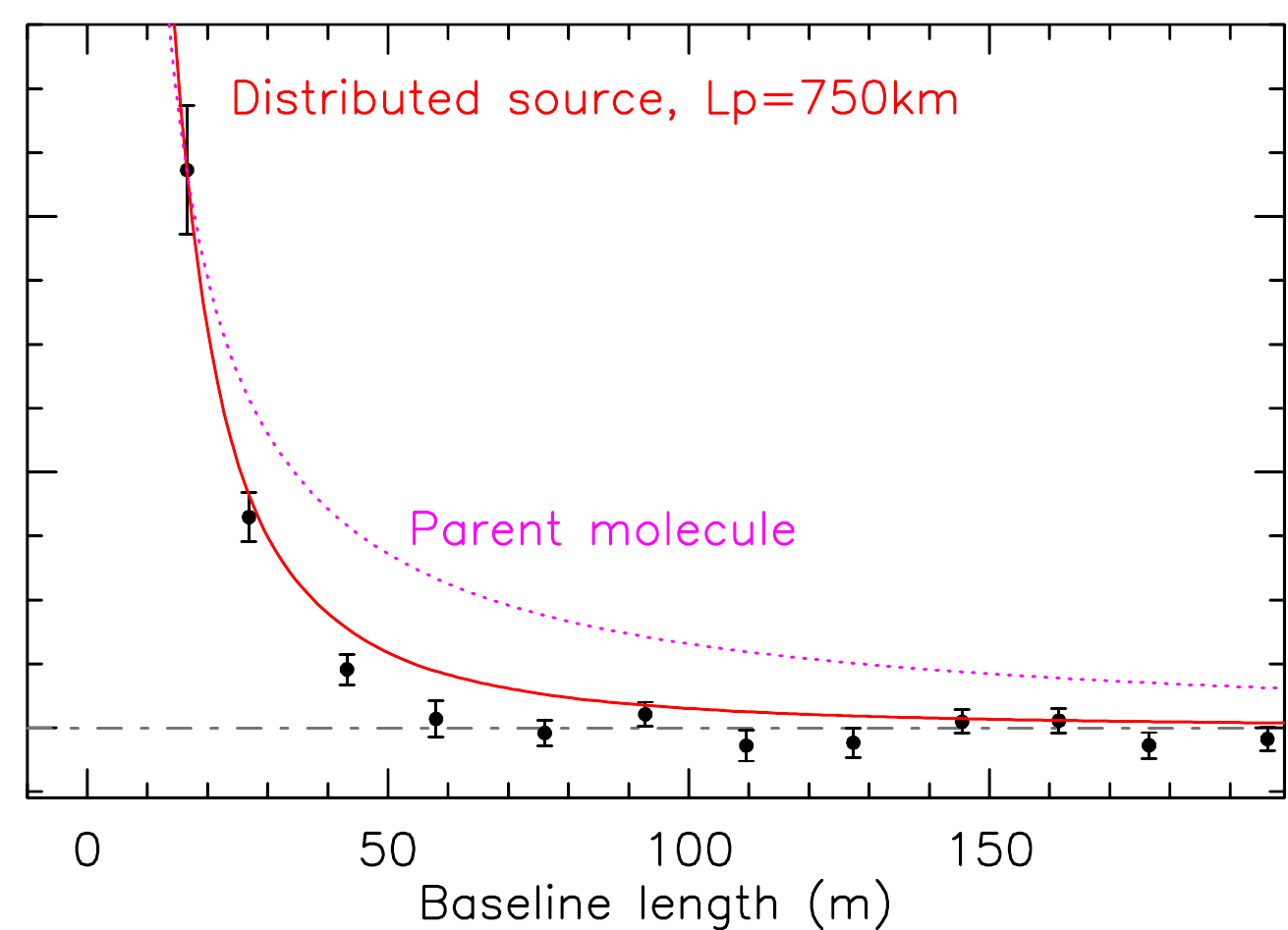}\\
\caption{Real part of the observed HNC ($4-3$) visibility amplitude ($Re$) \emph{vs.} baseline length, for each of the three time periods on Nov 17. { For the first two epochs, the best-fitting Haser daughter model (with $L_p=450$~km) is shown with a red curve. For the third epoch, only a lower limit for the HNC parent scale length ($L_p>750$~km) could be obtained.} Purple dotted curves show the simulated visibility amplitudes for a parent distribution, { which do not provide a good fit to the data}. \label{fig:vis}}
\end{figure*}

\subsection{2013 November 17}

By the start of our ALMA observations on Nov. 17 (approx. 26~hr later), the observed line and continuum fluxes had risen by about a factor of two relative to the start of observations on Nov. 16, and remained at a relatively constant level throughout the subsequent hour of ALMA observing. The total fluxes are shown as a function of time in the lower panel of Fig. \ref{fig:fluxvstime} and there is little evidence for any significant variation outside of the observational uncertainties.

The relative stability of the comet's activity is also apparent from the coma maps in Figs. \ref{fig:nov17_hnc} and \ref{fig:nov17_h2co}. Apart from the (random) noise features, the CH$_3$OH, H$_2$CO and continuum morphologies all show good consistency between the three time periods. The HNC maps, on the other hand, reveal a striking time variability that is most evident when comparing observation periods 1 and 3. In period 1, the HNC peak (detected at $>6\sigma$ level, where $\sigma$ is the RMS noise), appears at a position coincident with the continuum and other molecules, but in period 2, the HNC peak is shifted about 400~km towards the upper right of the map. Period 3 shows a weaker central emission peak and isolated, compact clump of strong ($6\sigma$) HNC emission (marked with an asterisk), at a position $870\pm25$~km north-west of the nominal nucleus position, at an angle $47^{\circ}$ clockwise from celestial north.

To highlight the significance of this temporal HNC variation, Fig. \ref{fig:profiles} shows one-dimensional slices through the HNC maps for each period, taken along the black dashed line shown in Fig. \ref{fig:nov17_hnc}, and passing through the center of the isolated clump. The shaded area in these plots represents the $1\sigma$ error envelope.  Note, in this analysis the signal-to-noise of our spectral profiles was insufficient to derive any useful radial velocity information for the gases, so our analysis is necessarily limited to the discussion of sky-projected distances in two dimensions, which is implicitly assumed from here on. 

In the first observation period, the 1-D HNC profile (Fig. \ref{fig:profiles}) is strongly centrally peaked. In period 2, the distribution is significantly broader and flatter, with a weakened central peak. In the third period the profile is broader and flatter still (with the central peak weakened to less than or about half its initial value), combined with the appearance of a spatially isolated HNC peak/clump far out into the coma. About 20\% of the total HNC flux is concentrated inside the $3\sigma$ contour surrounding this clump.

We followed the same methodology as \citet{cor14} to fit the visibility (Fourier) amplitude components of the observed HNC data using a Haser daughter model.  The real parts of the visibility amplitudes (visibilities) { were first rebinned to a uniform antenna separation (baseline) grid and are plotted as a function of baseline in Fig. \ref{fig:vis} for the three time periods. For each period, the visibilities show a relative excess of flux at large angular scales (short baselines) that cannot be explained if the observed HNC is a parent molecule, thus confirming the conclusions of \citet{lis08} and \citet{cor14} that HNC is a daughter/product species, with a distributed source in the coma.  For the first two time periods, the best-fitting HNC parent scale length ($L_p$) was found to be 450~km (with possible values in the range $L_p=250$-1200~km due to statistical uncertainties). For the third period, it was only possible to derive a lower limit from our data of $L_p>750$~km; this is due to an excess of extended flux that cannot be accounted for by our model. Our inability to fully constrain the scale of HNC production at that epoch suggests that the observed HNC distribution was even broader than can be explained using a standard Haser model, and further highlights the importance of coma production for this species.  For a more detailed interpretation of these data, a three-dimensional (anisotropic), time-variable radiative transfer model is required, which is beyond the scope of the present study, but is currently in development.}

\section{Discussion}

\subsection{Temporal variability of ISON's coma}
\label{sec:temporal}

Cometary activity is often observed to vary smoothly with heliocentric distance. Nevertheless, short time-scale variations (on periods of hours to days) are not uncommon \citep[\emph{e.g.}][]{mum86,sek02,li11,dra12}, and localized outbursts have recently been observed in detail by Rosetta in comet 67P/Churyumov-Gerasimenko \citep{gru16}. Our ALMA observations of C/2012 S1 (ISON), however, constitute the first detections of strong variability in HNC, H$_2$CO and sub-mm continuum over timescales of less than an hour. The drop in line and continuum fluxes by over 42\% in such a short timescale is remarkable as it implies an extremely abrupt, spontaneous change in the comet's activity. 

The observed rapid drop in line and continuum fluxes during our Nov. 16 observations can plausibly be explained by an abrupt decrease in the comet's gas and dust production near to the start of our ALMA observations.  Following this drop in activity, we hypothesize that the three main processes responsible for the loss of detected emission as a function of time are (1) molecular photodissociation, (2) sublimation of icy grains, and (3) transport of the coma material out of the ALMA field of view. The latter is expected due to the outflow/expansion of the coma (for gas and dust), with the additional possibility of differential sublimation pressure that may result in the acceleration of solid particles. 

To explore the possible impact on the observed fluxes of coma expansion following a reduction in comet activity, a simple, spherically-symmetric, constant-velocity outflow model was generated. Assuming the nucleus instantaneously stopped producing material at the start of our observations on Nov. 16 (at around UT 2013-11-16 10:10), and the coma was allowed to continue expanding outwards at the same velocity, the fraction of material remaining within the $2.5''$-radius integration aperture was calculated as a function of time and plotted with a dotted curve (for an outflow velocity $v_{out}=1.0$~\kms) and a dot-dashed curve (for $v_{out}=0.5$~\kms) in the upper panel of Fig. \ref{fig:fluxvstime}. These curves bracket the range of gradients in the observed line and continuum fluxes.

Given the reappearance of intense molecular and dust emission from the comet 26 hours later on Nov. 17, combined with the general trend for increasing Ly-$\alpha$ and optical flux on the days around our ALMA observations \citep[see][]{com14,sek14}, it seems unlikely that cometary activity ceased completely during our Nov. 16 observations. The decay in the ALMA fluxes was probably offset by at least some new gas and dust production. A more realistic scenario to explain our observations was obtained by constructing a time-dependent Haser daughter model for H$_2$CO, with fixed parent scale length of 280~km, H$_2$CO mixing ratio of 0.8\%, fixed temperature of $T=90$~K, $v_{out}=1.0$~\kms\ (see \citealt{cor14}) and a photodissociation rate of $6.7\times10^{-4}$~s$^{-1}$ (appropriate for $r_H=0.58$~AU; \citealt{hue92}). Synthetic images were generated using the CASA simulator, with the same observational and imaging characteristics as for our ALMA observations. Although our observations could be reproduced by many different time-variability scenarios, given our limited number of data points we adopt the simplest possible model to fit the data, in which the cometary H$_2$O production rate has an initial value $Q_{i}$, which falls (instantaneously) to a final value $Q_{f}$ after a time $t$. The best fit to our measured H$_2$CO fluxes is for $Q_{i}=1.05\times10^{29}$~s$^{-1}$, $Q_{f}/Q_i=0.49$ and $t=3.5$~min, which results in the decay curve shown with a dashed black line in Figure \ref{fig:fluxvstime}. Such a sharp $\approx50$\% reduction in the cometary production rate could be due, for example, to (1) the switching off of a powerful jet or large active region as a result of rotation-induced changes in the nucleus illumination, or (2) sudden depletion/exhaustion of volatiles from a component of the cometary material, which may be a consequence of the rapid evolution of a nucleus undergoing a series of disruptive outbursts.

Photodissociation can be ruled out as a loss process for the large dust grains thought to be responsible for the 0.8~mm continuum emission. However, if these grains contain a significant amount of ice, sublimation may be an important mechanism. For icy sub-mm grains at $r_H\approx0.5$~AU, \citet{bee06} calculated a lifetime (for complete sublimation) on the order of a few thousand seconds. The observed drop in continuum flux is therefore likely due to a reduction in dust grain production, followed by sublimation and the continued motion of the dust out of the ALMA field of view. Without sublimation, and assuming a constant size and temperature for the grains, a radial outflow velocity of $>0.5$~\kms\ is required to explain the more than a factor of two drop in continuum flux on Nov. 16 (Fig. \ref{fig:fluxvstime}, upper panel). Due to its small size, the thermal continuum emitted by the comet's nucleus is assumed to be negligible.

Due to a lack of previous mapping observations at mm/sub-mm wavelengths, the velocity distribution of large grains in cometary comae is not well established. \citet{sek12} determined a velocity of 30~m\,s$^{-1}$ for the submillimeter-sized dust released from sungrazing comet C/2013 W3 (Lovejoy) during its disruption. \citet{boi12} found grain outflow velocities up to 100~m\,s$^{-1}$ in the coma of 17P/Holmes, and during the Deep Impact event, $v_{out}\sim200$~m\,s$^{-1}$ was observed for the dust in 9P/Tempel 1 \citep{kel07}. Thus, the drop in continuum flux can probably not be explained by outflow alone (following a drop in the cometary activity), confirming the likely role of icy grain sublimation.  Our simplistic calculation does not account for the observed asymmetry (and implied anisotropic outflow) in the continuum maps. The properties of the large grains in comet ISON, including their density distribution, velocity field and sublimation rates may thus be better constrained through detailed (3-D) modeling of our ALMA continuum data in a future study.

\begin{figure*}
\centering
\includegraphics[width=0.32\textwidth]{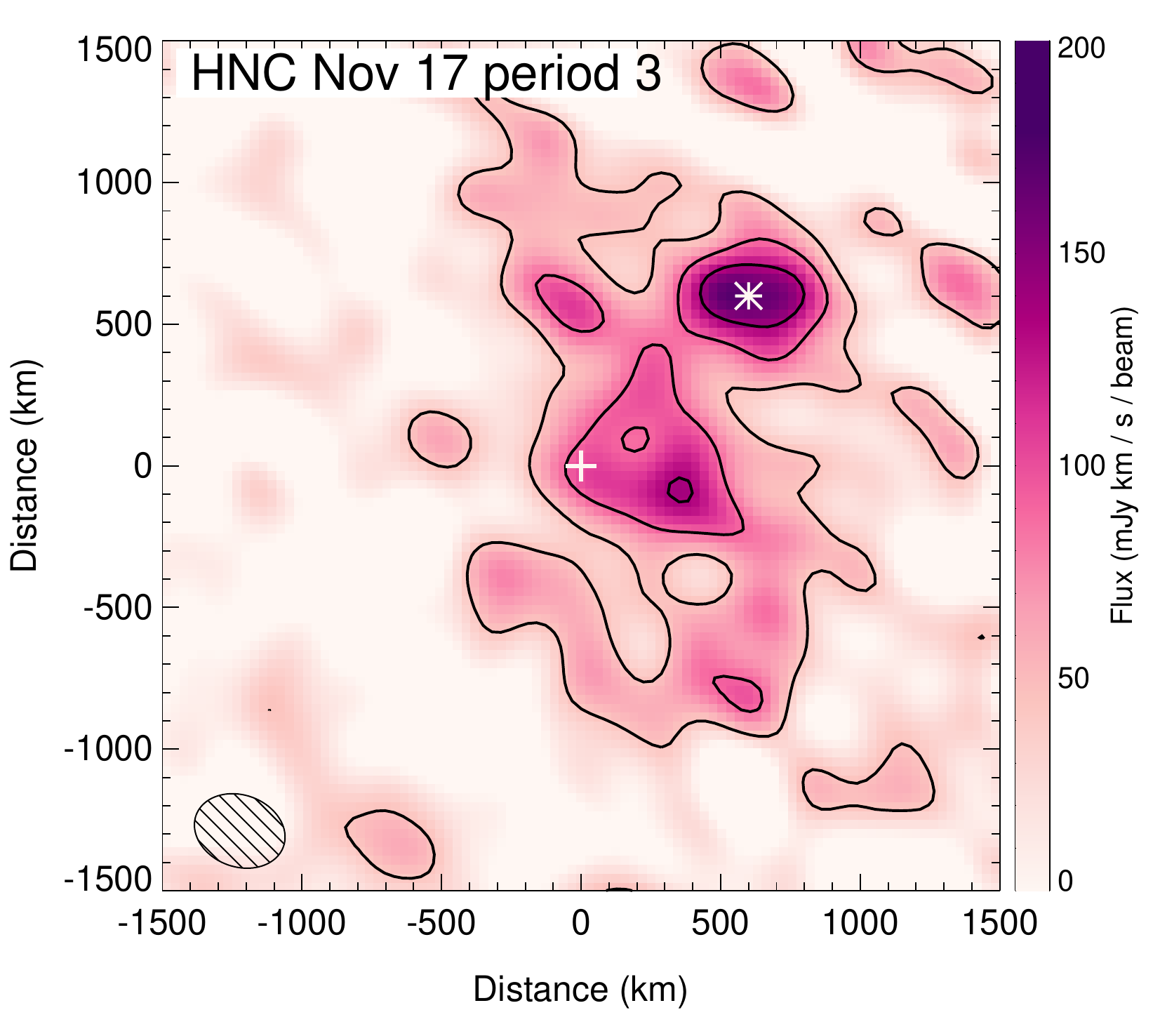}
\includegraphics[width=0.32\textwidth]{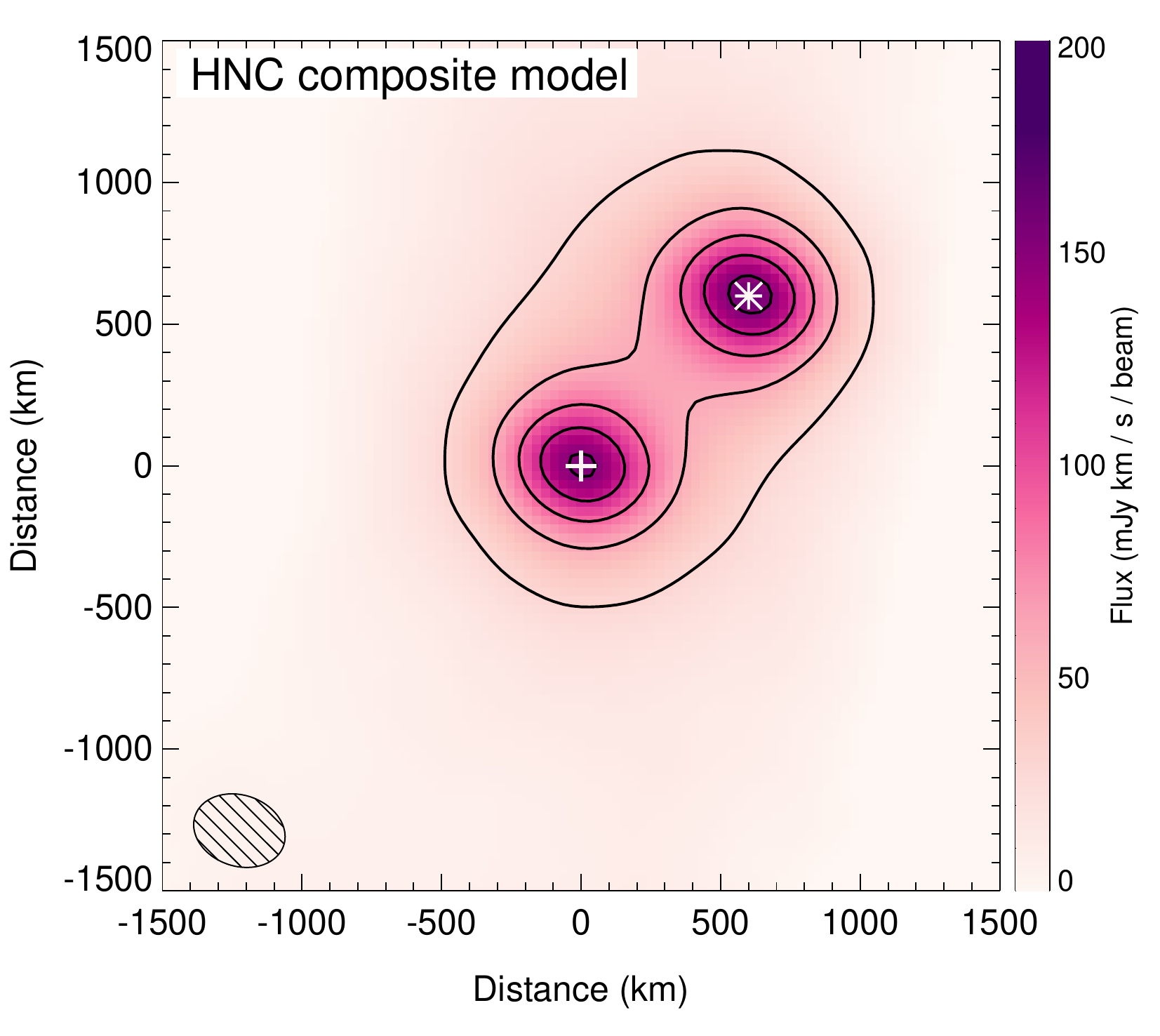}
\includegraphics[width=0.32\textwidth]{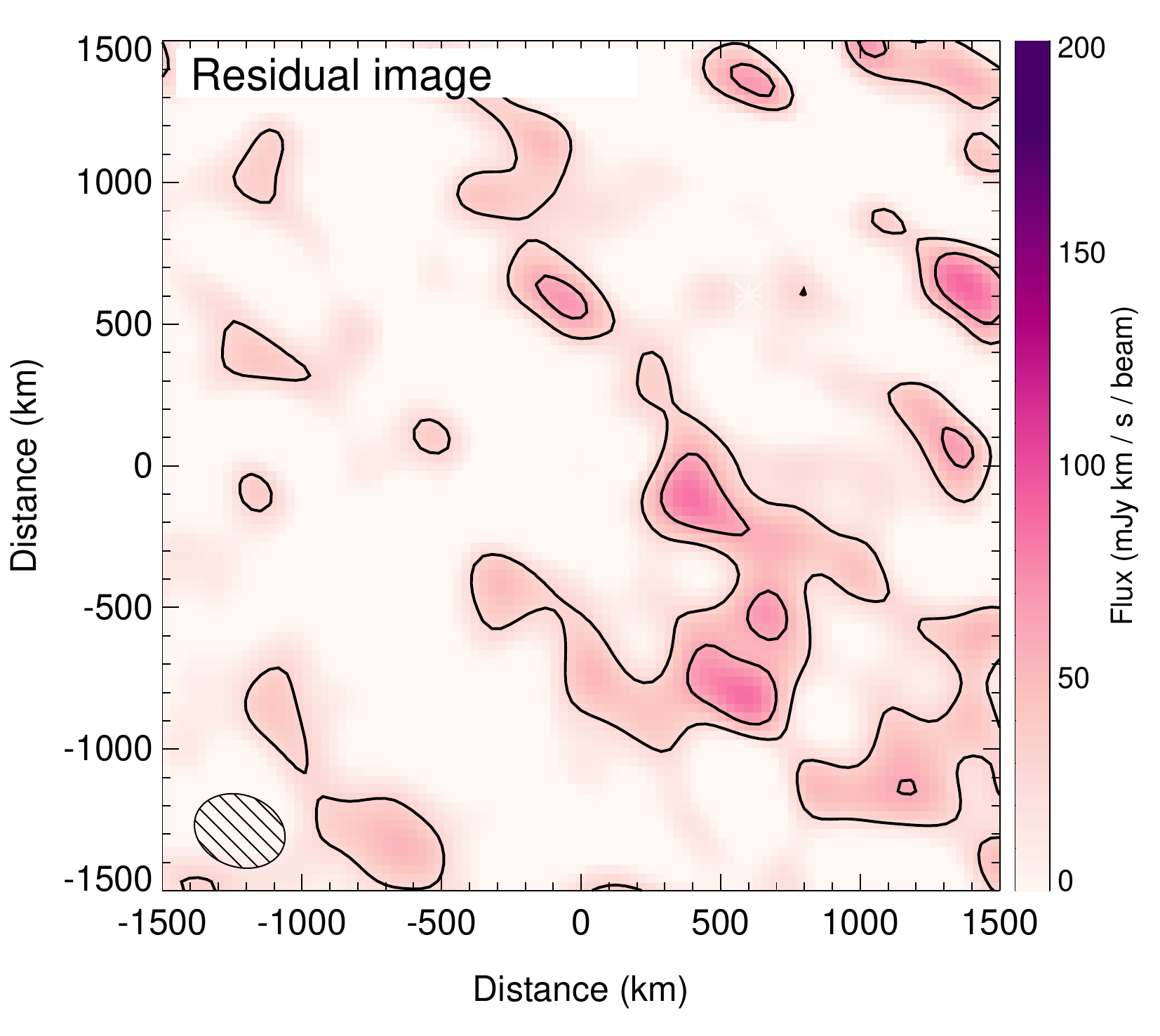}
\caption{Left: integrated flux map for HNC on Nov. 17, period 3. Middle: best-fitting two-component (composite) HNC production model. Right: residuals (model minus observed image); contour intervals are $1\sigma$.\label{fig:compmodel}}
\end{figure*}

Sublimation pressure of vaporizing material in comet ISON { (including water as well as more refractory species)} has been invoked to explain the anomalous non-gravitational motions experienced by the comet in its run up to perihelion \citep{sek14}, and such forces may also have played a role in the breakup of the nucleus \citep{ste15}. { Single-dish observations of ISON's sub-mm continuum emission by \citet{kea16} were found to be consistent with the non-gravitational model of \citet{sek14}, thus confirming the impact of intense sublimation pressure on the trajectory of the cometary material. The presence of a (time-variable) sub-mm emission tail in the anti-sunward direction (see Figures \ref{fig:nov16_h2co} and \ref{fig:nov17_h2co}), is consistent with this scenario.  We therefore conclude that the large grains responsible for the sub-mm continuum emission observed by ALMA} underwent enhanced heating on the sunward side, giving rise to a sublimation pressure that accelerated them away from the nucleus in the anti-sunward direction. {This acceleration was probably responsible for some of the drop in continuum flux observed on Nov. 16}.

\subsection{Elucidating the nature of cometary HNC release}

The time-evolution of the HNC distribution on Nov. 17 (Figs. \ref{fig:nov17_hnc} and \ref{fig:profiles}) seems to be most naturally interpreted as due to the release of an isolated clump of material from the nucleus that traveled radially outward through the coma at an average velocity of 0.41~\kms\ (in the plane of the sky), to reach a distance of 870~km in period 3. Given the HNC line FWHM of 2.0~\kms, the inferred clump outflow rate is reasonable (especially considering the possibility of an additional velocity component perpendicular to the plane of the sky), and is well within the 1~\kms\ outflow velocity for HCN and other species deduced by \citet{agu14} and \citet{cor14}. The derived outflow velocity is also consistent with the approximate limit of 0.5~\kms\ on the outflow rate of mm-sized dust. Such a transient, spatially isolated, directional outburst of material bears qualitative similarities to the collimated dust jets \citep{lin15} and decimeter-sized aggregates \citep{aga16}, emitted from confined regions on the surface of comet 67P/Churyumov-Gerasimenko, as observed by the Rosetta spacecraft.  

The production rate of HNC in the clump can be derived by modeling the flux map in period 3 of Nov. 17 using a two-component isotropic outflow model. A source of HNC was placed at the nominal nucleus position (white cross) and at the HNC peak (asterisk), both with $v_{out}=1$~\kms\ and $T=90$~K. Optimizing the HNC production rates of the two sources using a nonlinear least-squares procedure, an excellent fit to the observed image was obtained with $Q({\rm HNC})=2.9\times10^{25}$~s$^{-1}$ for the nucleus and $Q({\rm HNC})=3.2\times10^{25}$~s$^{-1}$ for the clump. A spectrally-integrated image of the best-fitting two-component HNC model is shown in the middle panel of Fig. \ref{fig:compmodel}, and the observation-minus-model image is shown in the right panel, for which the noise-like residuals indicate the quality of fit.

Given the expected $1/r^2$ falloff in the density of a parcel of expanding gas released from the nucleus, and the fact that the amplitude of the main HNC peak in Fig. \ref{fig:profiles} remains approximately constant despite traversing over a factor of two in distance between periods 2 and 3, it seems unlikely that such a well-defined clump could be released as gaseous HNC directly from the nucleus unless it was as an extremely narrow, well-confined HNC jet. Based on our interpretation of the HNC time series, a more likely explanation is that the observed HNC peak originates from an ice or organic dust-rich clump, aggregate or chunk of material emitted from the nucleus that underwent degradation/sublimation as it traveled out through the coma. If this material was mostly ice, then unless it was extremely HNC rich, the clump would also be expected to be apparent at the same positions in the H$_2$CO and CH$_3$OH maps. From the fact that the clump is not observed in H$_2$CO or CH$_3$OH but appears in the same general direction from the nucleus as the dust continuum tail, it follows that the clump likely originates from a macromolecular or polymeric (refractory) precursor more closely associated with the cometary dust than the ice. { The increase in relative HNC flux on large scales in the third epoch, as highlighted by the observed visibility amplitudes (Fig. \ref{fig:vis}), is also consistent with a spontaneous increase in the production of HNC in the coma as this clumpy material moved outward.}  

{
Our results are consistent with the hypothesis that cometary HNC originates from the degradation of macromolecules or polymeric material contained within the nucleus. Previously, \citet{rod01} and \citet{lis08} discussed possible macromolecular sources of cometary HNC, including hexamethylenetetramine (HMT; C$_6$H$_{12}$N$_4$), HCN polymer or other large organic molecules that undergo photolytic or thermal degradation as they travel out through the coma. Refractory organic particles are known to exist in comets \citep{cot04}, and are believed to have originated in the interstellar medium or protosolar disk. Polyoxymethylene (POM), in particular, has been studied in detail as a candidate to explain the distributed H$_2$CO sources in comets 1P/Halley and C/1995 O1 (Hale-Bopp) \citep{cot04,fra06}. Recent in-situ work on comet 67P using the Rosetta COSIMA instrument revealed the presence of very large macromolecular compounds in the cometary dust, analogous to the insoluble organic matter found in carbonaceous meteorites \citep{fra16}. Possible evidence for a radiation-induced, POM-like polymer was also found on the surface of 67P by the Ptolemy instrument \citep{wri15}, but a conclusive identification of specific macromolecules (including HMT, POM and HNC polymer), remains elusive (C. Altwegg, private communication 2016). Now that detailed mapping of cometary molecular distributions has become routine (thanks to ALMA), additional comparisons between laboratory data and astronomical observations on the breakdown of large molecules will be possible, to further constrain the nature of cometary organic material.
}

Based on the differences between the HNC and H$_2$CO coma morphologies, we conclude that the organic dust source for HNC is not related to that responsible for the distributed source of cometary H$_2$CO. This implies the presence of at least two kinds of macromolecular parent material in comets: nitrogen-rich and oxygen-rich, the identification of which remains a challenge for laboratory astrophysics.

\section{Conclusion}

Spatially and temporally resolved observations of molecular and dust emission from C/2012 S1 (ISON) were obtained as the comet passed between 0.58 and 0.54~AU of the sun on its approach to perihelion in November 2013. The observed maps reveal a complex and dynamic coma, with molecular line and continuum features evolving rapidly over periods of tens of minutes. The anisotropic HNC outflow features detected by \citet{cor14} are found to be due to the presence of time-variable HNC clumps or well-confined jets moving outward through the coma, and may be explained as the result of the breakdown of HNC-rich icy particles, or refractory materials including organic dust grains, polymers or other macro molecules. The exact nature of the HNC parent has been debated in the literature for some time \citep{irv98,rod05,lis08,cor14}, but our observations strongly suggest an origin for cometary HNC in organic refractory material, released intermittently from confined regions on the nucleus. 

A rapid decay in the observed gas emission on November 16 (over a period of 50 minutes) is probably the result of a substantial reduction in the activity of the comet during our observations, followed by partial dissipation of the coma through photodissociation and outflow processes. Using a Haser daughter model the H$_2$CO observations on Nov 16 can be explained as being due to a $\approx50$\% drop in the cometary production rate near the beginning of our observations. The surprisingly rapid decay in mm continuum flux implies rapid outflow, probably combined with the partial sublimation of large, icy grains. Following this lull, coma activity had returned strongly by UT 12:30 on November 17, about a factor of two higher than at the start of our observations on Nov. 16. 

Combined with previous radio and optical observations, these data reveal an extremely erratic behavior in the outgassing of volatiles from comet ISON as it approached the Sun, with periods of strong outburst interspersed with quiescent intervals, resulting in sporadic production of gas and dust. Future studies of cometary outgassing and volatile depletion during nucleus disruption will be required to help explain the origin of this behavior. The surprising coma variability detected with ALMA demonstrates the importance of temporally and spatially-resolved observations on the smallest scales in order to properly understand the structure and composition of cometary comae. { Detailed maps of the distributions of coma product species such as HNC and H$_2$CO, combined with laboratory experiments on the breakdown of candidate parent materials will be required to obtain a full understanding of the origins and complex behaviors of these intriguing cometary species}.

\acknowledgments
This study was supported by the National Science Foundation under Grant No. AST-1616306, by NASA's Planetary Atmospheres and Planetary Astronomy Programs and by the NASA Astrobiology Institute through the Goddard Center for Astrobiology. It makes use of ALMA data set \#2012.A.00033.S. ALMA is a partnership of ESO (representing its member states), NSF (USA) and NINS (Japan), together with NRC (Canada) and NSC and ASIAA (Taiwan), in cooperation with the Republic of Chile. The Joint ALMA Observatory is operated by ESO, AUI/NRAO and NAOJ. The National Radio Astronomy Observatory is a facility of the National Science Foundation operated under cooperative agreement by Associated Universities, Inc. DCL is supported by NASA through JPL/Caltech. DM is supported by Basal CATA PFB-06 and the ICM MAS. YJK is supported by NSC grants 99-2112-M-003-003-MY3 and 100-2119-M-003-001-MY3.

\end{document}